%% file: main.tex
\documentclass[sigconf,nonacm]{acmart}

\usepackage{enumitem}
\usepackage{hyperref}
\usepackage{multirow}
\usepackage{amsthm}
\setlist[enumerate]{align=left}
\usepackage{algorithm}
\usepackage{algorithmic}
\usepackage{caption}
\usepackage{subcaption}
\usepackage{bm}
\usepackage{xcolor}
\usepackage{colortbl}
\definecolor{marron}{HTML}{CC88B0}
 \usepackage{amsmath,amsthm,mathtools, bm}

\usepackage[multiple]{footmisc}

\AtBeginDocument{%
  }

\copyrightyear{2027}
\acmYear{2027}
\setcopyright{none}
\acmDOI{XXXXXXX.XXXXXXX}
\acmISBN{978-1-4503-XXXX-X/2018/06}

\begin{document}

\title{Towards a Unified Framework of Optimal Input-Side Dimensional Configuration for Deep CTR Prediction}

\author{Chaohua Yang}
\affiliation{%
  \institution{College of Computer Science and Software Engineering, Shenzhen University}
  \city{Shenzhen}
  \country{China}
}
\email{chaohua.ych@gmail.com}

\author{Fuyuan Lyu}
\affiliation{
  \institution{McGill University}
  \city{Montreal}
  \country{Canada}
}
\email{fuyuan.lyu@mail.mcgill.ca}

\author{Xing Tang}
\affiliation{
  \institution{Shenzhen Technology University}
  \city{Shenzhen}
  \country{China}
}
\email{xing.tang@hotmail.com}

\author{Dugang Liu}
\affiliation{%
  \institution{College of Computer Science and Software Engineering, Shenzhen University}
  \city{Shenzhen}
  \country{China}}  
\email{dugang.ldg@gmail.com}

\author{Xiuqiang He}
\affiliation{%
  \institution{Shenzhen Technology University}
  \city{Shenzhen}
  \country{China}
}
\email{hexiuqiang@sztu.edu.cn}

\author{Zhong Ming}
\affiliation{%
  \institution{College of Computer Science and Software Engineering, Shenzhen University}
  \city{Shenzhen}
  \country{China}
}
\email{mingz@szu.edu.cn}

\renewcommand{\shortauthors}{Chaohua Yang, et al.}

\input{sec/0_abstract}



\maketitle

\input{sec/1_introduction}
\input{sec/2_related_work}
\input{sec/3_preliminaries}
\input{sec/4_methodology}
\input{sec/5_experiments}
\input{sec/6_conclusion}

\bibliographystyle{ACM-Reference-Format}
\bibliography{sample-base}

\appendix
\input{sec/7_appendix}

\end{document}

%% file: sec/0_abstract.tex
\begin{abstract}
Deep click-through rate (CTR) prediction has achieved substantial success through increasingly expressive feature interaction architectures.
However, predictive performance also depends on how input-side information is configured before interaction modeling. Existing studies have shown that selecting informative samples, features, or embedding units can improve prediction accuracy. 
Nevertheless, they primarily optimize these selection mechanisms independently along a single input-side dimension, overlooking their shared basic form and the inherent dependencies across input levels, thereby limiting the effectiveness of input-side dimensional configuration.
To address this issue, we propose OptIDC, a unified framework for Optimal Input-side Dimensional Configuration in deep CTR prediction. OptIDC jointly learns sample, feature, and embedding-unit level configurations within a unified framework by integrating their corresponding selection mechanisms. 
To explicitly coordinate selection decisions across different levels, OptIDC further introduces a selection-weighted Optimal Transport mechanism that aligns coarse- and fine-grained selections via transport over finite measures, thereby alleviating cross-level selection conflicts while preserving the semantics of each input level. 
Moreover, OptIDC is model-agnostic and can be seamlessly integrated into existing CTR models.
Extensive experiments on four real-world benchmark datasets demonstrate that OptIDC consistently achieves higher AUC than existing input-side selection methods, validating the effectiveness of unified input-side dimensional configuration.
\end{abstract}

%% file: sec/1_introduction.tex
\section{Introduction}
\label{sec:introduction}
In web applications such as online advertising and recommendation systems, click-through rate (CTR) prediction estimates the probability that a user will click on a candidate item if displayed and supports content ranking.
Much of the progress in deep CTR prediction has come from increasingly expressive architectures for modeling feature interactions~\cite{cheng2016wide,zhang2019deep}.
These architectures, however, operate on the information supplied to them. Real-world training data contain samples of uneven quality, feature fields of varying relevance, and embedding units that do not contribute equally to prediction. 
CTR prediction performance therefore depends on two complementary choices: how information is modeled by the architecture and what information is allowed to shape the model. 
We use the term \emph{input-side dimensional configuration} for the latter, spanning the sample, feature, and embedding-unit levels.

\begin{figure}[htbp]
    \centering
    \includegraphics[width=0.48\textwidth]{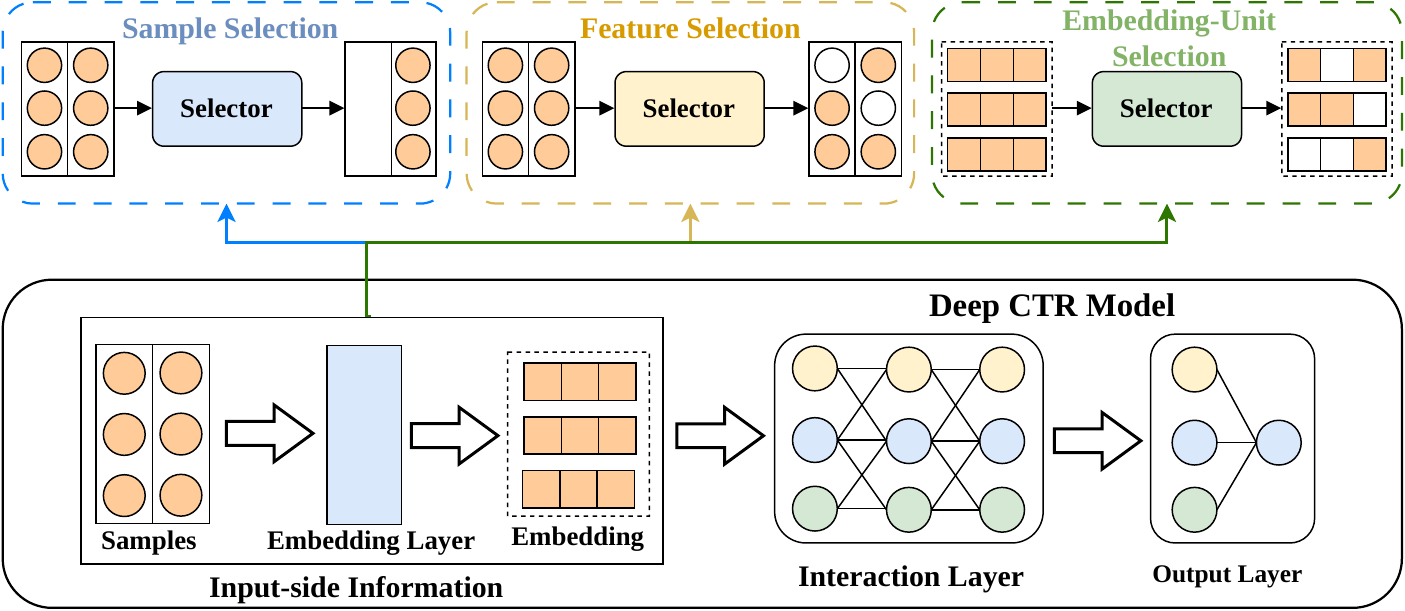}
    \Description[<Figure 1. Fully described in the text.>]{<A full description of Figure 1 can be found in the second paragraph of Section 1.>}
    \caption{Overview of the input-side selection methods for deep CTR prediction.}
    \label{fig:overview}
    \vspace{-10pt}
\end{figure}

Fig.~\ref{fig:overview} places existing input-side selection methods within a common deep CTR pipeline.
At the sample level, instance selection and data denoising reweight or remove training examples~\cite{gao2023autotransfer,thanthriwatta2021instance, Autodenoise}. At the feature level, field- and value-level selection methods learn gates that retain informative inputs before interaction modeling~\cite{lin2022adafs, autofield, optfs}. At the embedding-unit level, pruning and dimension search allocate representation capacity at a finer granularity~\cite{liu2020automated, AutoDim, optembed, PEP}. 
The gains reported across this substantial body of work confirm that input configuration can materially affect CTR prediction performance and efficiency. 
In most prior work, however, only one of these three selection paths is optimized at a time, while the other input levels are treated as fixed. Decisions that ultimately shape the same model input are therefore learned without explicit coordination across levels.

Despite their different names and granularities, the three selection mechanisms share the same basic form: each assigns a selection weight to an input component and uses that weight to control the component's contribution to model learning. Sample selection assigns one weight to an entire instance, feature selection assigns weights to its field representations, and embedding-unit selection assigns weights to individual coordinates within those representations. When expressed over the embedding tensor, they can all be viewed as masks applied along different input axes. This common form also exposes their dependence. A sample contributes through its selected feature embeddings, and a feature contributes through its selected embedding units. If a feature is masked out, any unit-level decisions within it become ineffective; if nearly all feature representations of a highly weighted sample are suppressed, the sample-level decision has little effect. We therefore view the three mechanisms as coupled decisions about the same effective input, rather than as three unrelated selection problems.

Their common form motivates joint modeling. Blindly attaching all three selection mechanisms to the same backbone, however, leaves the final prediction error as their only cross-level signal. Sample weighting controls how much gradient an instance contributes, feature gating controls which field representations receive that gradient, and embedding-unit masking controls which coordinates carry it. Their updates can therefore interfere: a low sample weight may deprive the finer-grained mechanisms of useful feedback, while aggressive feature or embedding-unit masks can make a highly weighted sample almost ineffective. Effective joint configuration must preserve useful specialization at each level and control severe cross-level mismatches, which we call \emph{selection conflicts}. It must also coordinate outputs with different structures: a sample weight is a scalar, a feature gate is defined over fields, and an embedding-unit mask is defined over individual coordinates. These requirements expose the central challenge of unified input-side configuration: \emph{how can heterogeneous selection mechanisms be explicitly coordinated while preserving the granularity and semantics of each input level?}

To tackle this challenge, we propose \textbf{OptIDC} (\textbf{Opt}imal \textbf{I}nput-side \textbf{D}imensional \textbf{C}onfiguration), a unified framework that learns configurations at the sample, feature, and embedding-unit levels. OptIDC introduces a selection-weighted optimal transport (OT) formulation over finite measures for cross-level coordination~\cite{cuturi2013sinkhorn}. For each coarse-to-fine relation, the coarse-level selection weight determines the total transported mass, the finer-grained weights allocate this mass over their semantic support, and representation dissimilarity defines the transport cost. This formulation accommodates different selection cardinalities and retains the semantics of each level through the representation-based cost. We construct three OT objectives for the sample--feature, feature--embedding-unit, and sample--embedding-unit relations. Together, they provide direct coordination signals between adjacent and non-adjacent levels, encouraging each coarse-grained decision to be supported by compatible fine-grained decisions.
During configuration learning, the sample weights, feature gates, embedding-unit masks, and transport costs are optimized together with the prediction objective. The learned configuration is then used to retrain the backbone with the selected input information. Since OptIDC operates on the input side, it leaves the backbone's interaction architecture unchanged and can be used with different CTR models.
We evaluate the framework on four public datasets and five representative backbones. Across the evaluated settings, joint configuration improves over the corresponding single-level alternatives, and OT coordination provides further gains over joint configuration alone. Ablation, efficiency, and mask analyses further examine the source of these gains.

Our main contributions are summarized as follows:
\begin{itemize}[leftmargin=*]
    \item We connect sample, feature, and embedding-unit selection through their common masking form and formulate the input-side dimensional configuration as a coupled problem over the three input levels.
    \item We develop OptIDC with a selection-weighted coordination objective for the sample--feature, feature--embedding-unit, and sample--embedding-unit relations, combining coarse selection mass, fine-level allocation, and representation compatibility through a finite-measure OT formulation.
    \item We evaluate OptIDC on four datasets and five CTR prediction backbones. The experiments compare it with single-level alternatives and disentangle the effects of joint configuration and OT coordination through ablation, efficiency, and selection consistency analyses.
\end{itemize}

%% file: sec/2_related_work.tex
\section{Related Work}\label{sec:related}
In this section, we briefly review some relevant works on two research topics, including deep CTR prediction and automated selection in CTR prediction.

\noindent\textbf{Deep CTR Prediction.}
Deep learning has become a widely adopted approach to CTR prediction due to its strong capacity for modeling complex interactions among user, item, and contextual features.
Early models, such as factorization machines, model pairwise interactions among sparse input features~\cite{rendle2010fm}, while deep architectures such as Wide\&Deep and DeepFM further capture higher-order interactions~\cite{cheng2016wide, DeepFM}.
Subsequent advances further improve modeling capacity through explicit feature-interaction networks, attention mechanisms, and multi-task learning~\cite{wang2017deep, song2019autoint, ma2018modeling, ple}.
Despite their success, these architectural approaches typically treat all input-side information, such as samples, features, and embedding units, equally, without selecting or weighting informative signals during modeling.
Thus, redundant or uninformative inputs may still limit efficiency and predictive performance when handling large-scale and high-dimensional data.
In contrast, OptIDC complements these architectural advances by jointly configuring information at the sample, feature, and embedding-unit levels for CTR prediction.

\noindent\textbf{Automated Selection in CTR Prediction.}
Introducing automated selection mechanisms into CTR prediction has attracted increasing attention in recent years as an effective way to improve prediction performance~\cite{zheng2023automl}.
Existing studies mainly investigate information selection in three input-side dimensions.
For feature inputs, prior work focuses on identifying informative feature subsets through field-level or value-level selection to reduce redundancy~\cite{lin2022adafs, autofield, optfs}.
For embedding representations, embedding-unit or -dimension selection methods further refine representations by retaining important embedding elements and reducing unnecessary parameters~\cite{AutoDim, liu2020automated, PEP}.
OptEmbed~\cite{optembed} combines feature selection with embedding-dimension search.
For training instances, sample selection methods identify high-value or reliable samples for model training~\cite{thanthriwatta2021instance, Autodenoise, gao2023autotransfer}.
Beyond input-side information, automated selection has also been explored for feature interaction~\cite{lyu2023towards, liu2020autofis}, loss functions~\cite{zhao2021autoloss}, and hyperparameters~\cite{chen2019lambdaopt}.
Unlike these methods, which usually optimize one selection space at a time, our work extends input-side selection to a unified setting that jointly considers the sample, feature, and embedding-unit levels.

%% file: sec/3_preliminaries.tex
\section{Preliminaries}
\label{sec:preliminaries}

\subsection{Motivating Experiments}
\label{sec:motivating-experiments}

\begin{figure}[htbp]
\vspace{-5pt}
\centering
\subfloat[The performance of different  selection methods.]{\includegraphics[width=0.44\linewidth]{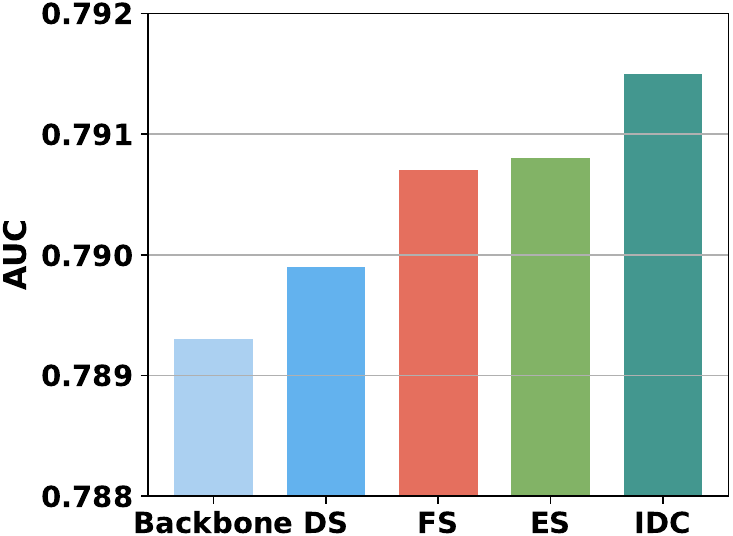}
\label{fig:motivating-case1}}
\hspace{0.1em}
\subfloat[The correlation between selection consistency and performance.]{\includegraphics[width=0.5\linewidth]{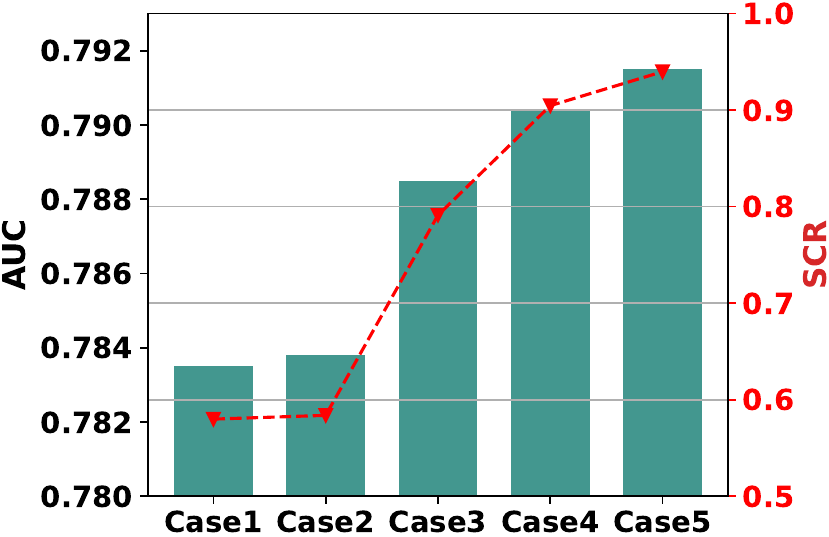}
\label{fig:motivating-case2}} 
\Description[<Figure 2. Fully described in the text.>]{<A full description of Figure 2 can be found in Section 3.1.>}
\caption{(a) A comparison of the performance of different selection methods on the Avazu dataset. (b) The association between selection consistency and model performance on the Avazu dataset.}
\vspace{-10pt}
\end{figure}

We first examine whether the three input levels benefit from joint configuration. Using Avazu with the DNN as the backbone, we compare sample selection (DS), feature selection (FS), and embedding-unit selection (ES) with \emph{Input-side Dimensional Configuration} (IDC), which learns all three selectors under the CTR prediction objective without explicit cross-level coordination. As shown in Fig.~\ref{fig:motivating-case1}, the single-level variants improve AUC by about 0.001 over the backbone, whereas IDC yields a gain of more than 0.002. This result suggests that the three levels provide complementary information when configured together.

Joint learning, however, leaves the prediction loss as the only signal shared by selectors of different granularities. We therefore examine five IDC cases with different cross-level selection patterns. After mapping their binary decisions to a common space (see the Appendix~\ref{sec:mask mapping mechanism} for the mapping process), and inspired by the Jaccard similarity, we measure the overlap among the three masks using:
\begin{equation}
\label{eq:three-way-scr}
\text{SCR}(\mathbf{M}^\mathrm{s}, \mathbf{M}^\mathrm{f}, \mathbf{M}^\mathrm{e}) =
\frac{
\left\lVert \min(\mathbf{M}^\mathrm{s},\mathbf{M}^\mathrm{f},\mathbf{M}^\mathrm{e}) \right\rVert_1
}{
\left\lVert \max(\mathbf{M}^\mathrm{s},\mathbf{M}^\mathrm{f}, \mathbf{M}^\mathrm{e}) \right\rVert_1
},
\end{equation}
where $\left\lVert \cdot \right\rVert_1$ denotes the $l_1$ norm, $\max(\cdot)$ and $\min(\cdot)$ represent the element-wise maximum and minimum operations, respectively.
The SCR metric remains bounded within [0,1], where a larger SCR indicates greater
agreement in both support and mapped activation magnitude.
Fig.~\ref{fig:motivating-case2} shows that SCR rises by about 0.30 across these cases, accompanied by an AUC gain of nearly 0.008. Although this association is not causal evidence, it motivates explicit coordination rather than relying on a shared prediction loss alone.

\subsection{Problem Formulation}
\label{sec:problem-formulation}

Let $\mathcal D=\{(\mathbf x_i,y_i)\}_{i=1}^{N}$ be a CTR prediction dataset, where $\mathbf x_i$ contains $F$ feature fields and $y_i \in \{0,1\}$ is the target, where $y_i =1$ denotes \textit{click}. An embedding layer maps $\mathbf x_i$ to
$\mathbf H_i=[\mathbf h_{i1},\ldots,\mathbf h_{iF}]^\top\in\mathbb R^{F\times d}$, where $d$ is the embedding dimension. 
Then, $\mathbf H_i$ is consumed by a backbone model $f_{\theta}$. 
Based on the definition above, we associate each sample with a sample weight $s_i\in[0,1]$, a feature gate $\mathbf g_i\in[0,1]^F$, and an embedding-unit mask $\mathbf M_i\in[0,1]^{F\times d}$. Their selection scores are continuous during configuration learning and can be discretized to obtain the final configuration.

Although these have different shapes, they perform the same operation at different resolutions. The feature and unit decisions determine what enters the backbone model, while the sample decision determines how strongly that input contributes to learning.
The two fine-grained masks first produce the
configured embedding, represented as follows:
\begin{equation}
\label{eq:configured-embedding-problem}
\widetilde{\mathbf H}_i = \mathbf H_i \odot(\mathbf g_i\mathbf 1_d^\top) \odot\mathbf M_i,
\end{equation}
Here $\odot$ denotes element-wise multiplication, and $\mathbf{1}_d \in \{1\}^d$ is an all-one vector. The sample weight then weights the loss evaluated on $\widetilde{\mathbf H}_i$ from
Eq.~\eqref{eq:configured-embedding-problem}, which can be formalized as:
\begin{equation}
\label{eq:unified-recommendation}
    \mathcal L_{\mathrm{ctr}}
    =
    \frac{1}{N}\sum_{i=1}^{N}
    s_i \cdot\,
    \ell\!\left(
        y_i,f_{\theta}(\widetilde{\mathbf H}_i)
    \right),
\end{equation}
where $\ell(\cdot)$ is the cross-entropy loss function.
Optimizing Eq.~\eqref{eq:unified-recommendation} with separate selection regularizers yields IDC. The selectors share the prediction loss, but the objective contains no term that directly relates a coarse decision to the finer-grained decisions through which it takes effect.

Thus, we consider the sample--feature, feature--embedding-unit, and sample--embedding-unit relations, abbreviated as $\mathrm{sf}$, $\mathrm{fe}$, and $\mathrm{se}$. 
For sample $i$, let
$\mathcal W_{\mathrm{sf}}(i)$,
$\mathcal W_{\mathrm{fe}}(i)$, and
$\mathcal W_{\mathrm{se}}(i)$ denote differentiable costs that measure whether a coarse decision is supported by compatible fine-grained components. Their dependence on the selection outputs and learned representations is suppressed here for readability. 
The coordination target is defined as follows:
\begin{equation}
    \mathcal L_{\mathrm{coord}}
    =
    \frac{\lambda_{\mathrm{co}}}{N}
    \sum_{i=1}^{N} 
    \left[
        \mathcal W_{\mathrm{sf}}(i)
        +
        \mathcal W_{\mathrm{fe}}(i)
        +
        \mathcal W_{\mathrm{se}}(i)
    \right],
    \label{eq:coordination-target}
\end{equation}
where the nonnegative $\lambda_{\mathrm{co}}$ coefficient controls the relations.
Finally, the optimal input-side dimensional configuration problem combines CTR prediction, level-specific selection constraints, and cross-level coordination as follows:
\begin{equation}
    \min_{\theta, \Theta, \boldsymbol\varphi}
    \mathcal L_{\mathrm{ctr}}
    +\Omega_{\mathrm{sel}}(\mathbf s,\mathbf G,\mathbf M)
    +\mathcal L_{\mathrm{coord}},
    \label{eq:optidc-problem}
\end{equation}
where $\Theta$ is the embedding parameters, $\theta$ is the backbone model parameters, $\mathbf s=(s_i)_{i=1}^{N}$, $\mathbf G=(\mathbf g_i)_{i=1}^{N}$, $\mathbf M=(\mathbf M_i)_{i=1}^{N}$ are the outputs induced by
$\boldsymbol\varphi$, and $\Omega_{\mathrm{sel}}$ controls the desired
amount of selection at each level. 
Eq.~\eqref{eq:optidc-problem} states the desired behavior but leaves two choices open: how to produce the three selectors along one information path, and how to instantiate the three costs without forcing heterogeneous decisions onto an artificial common support. The next section addresses these two choices.

%% file: sec/4_methodology.tex
\section{The OptIDC Framework}
\label{sec:methodology}

\subsection{Framework Overview}
\label{sec:framework-overview}
\begin{figure*}[htb]
    \centering
    \includegraphics[width=1.\textwidth]{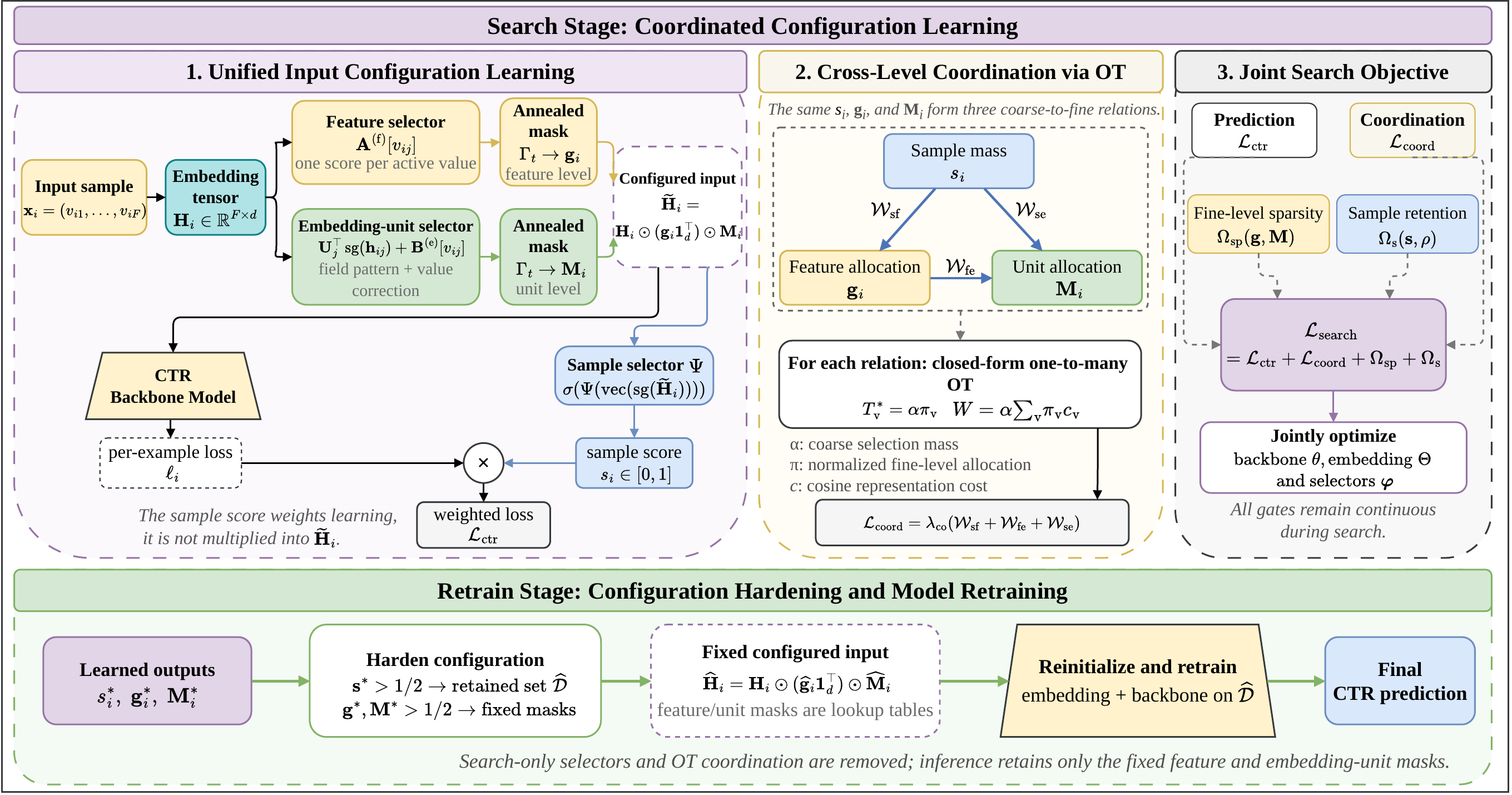}
    \Description[<Figure 3. Fully described in the text.>]{<A full description of Figure 3 can be found in Section 4.1.>}
    \caption{The architecture of the Optimal Input-side Dimensional Configuration (OptIDC) framework.}
    \label{fig:framework}
    \vspace{-10pt}
\end{figure*}

\textbf{OptIDC} is organized around two questions raised by unified input configuration: how to learn sample, feature, and embedding-unit decisions within the same model, and how to keep these decisions from evolving independently. 
For the first question, we arrange the three selectors along the path by which an input reaches the CTR prediction model. Feature and embedding-unit decisions determine what the backbone observes, and the sample decision evaluates how strongly that configured input should influence learning. 
For the second, we connect every coarse-to-fine pair through a selection-weighted optimal transport (OT) cost that respects the native granularity of both selectors.
The framework follows a search-and-retrain workflow. Continuous gates make configuration choices differentiable during search, allowing the selectors and CTR prediction backbone to be learned together. Once a configuration is identified, the gates are converted into hard decisions, and the backbone is retrained with this configuration fixed. 
The overall architecture of OptIDC is illustrated in Fig.~\ref{fig:framework}.
The following subsections describe the unified input configuration learning, its cross-level coordination, and the two-stage optimization.

\subsection{Unified Input Configuration Learning}
\label{sec:unified-configuration}
The formulation in Eq.\eqref{eq:configured-embedding-problem} and Eq.\eqref{eq:unified-recommendation} tells us how the three decisions act, but not how they are produced. OptIDC learns them in the same order in which they affect the input. Feature and embedding-unit selectors first configure the embedding tensor. The sample selector then scores the resulting tensor, so its decision is based on the information that the backbone actually receives.

For sample $i$, let $v_{ij}=x_{ij}$ be the active feature value in field $j$, and write $[n]=\{1,\ldots,n\}$. Let $\mathcal V$ be the disjoint union of the field vocabularies.
A lookup table $\mathbf A^{(\mathrm f)}\in\mathbb R^{|\mathcal V|}$ provides one feature logit for each value, where $n = |\mathcal V|$ is the total number of feature values.
To score individual embedding units, field $j$ owns unit representations $\mathbf U_j=[\mathbf u_{j1},\ldots,\mathbf u_{jd}] \in\mathbb R^{d\times d}$, and $\mathbf B^{(\mathrm e)}\in\mathbb R^{|\mathcal V|\times d}$ provides a value-specific bias. Their two logits are expressed as:
\begin{equation}
    a_{ij}^{(\mathrm f)}
    =
    \mathbf A^{(\mathrm f)}[v_{ij}],
    \qquad
    \mathbf a_{ij}^{(\mathrm e)}
    =
    \mathbf U_j^\top
    \operatorname{sg}(\mathbf h_{ij})
    +
    \mathbf B^{(\mathrm e)}[v_{ij}],
    \label{eq:selector-logits}
\end{equation}
where the first equality retrieves a scalar score for the active feature value. 
Because this score controls its entire $d$-dimensional embedding, it answers a feature-level question: whether $v_{ij}$ should contribute as a whole. 
The second equality asks the finer question of which units should contribute. Component-wise, its $k$-th logit is $a_{ijk}^{(\mathrm e)} =\mathbf u_{jk}^{\top}\operatorname{sg}(\mathbf h_{ij}) +\mathbf B^{(\mathrm e)}[v_{ij},k]$. 
The inner product measures the compatibility between the current embedding and unit representation $\mathbf u_{jk}$. Since $\mathbf U_j$ is shared by all values in field $j$, it captures a reusable field-level scoring pattern, while $\mathbf B^{(\mathrm e)}[v_{ij},k]$ provides a value-specific correction. The stop-gradient operator $\operatorname{sg}(\cdot)$ prevents the unit-scoring branch from reshaping the embedding merely to make its selection easier.
Then, following~\cite{optfs, MultiFS}, the logits are converted to differentiable gates with a shared annealing schedule:
\begin{equation}
\label{eq:embedding-configuration-gates}
        \Gamma_t(a) = \sigma\!\left(\gamma^{\,t/T}a\right),
        \quad
        g_{ij} = \Gamma_t(a_{ij}^{(\mathrm f)}),
        \quad
        \mathbf M_{ij:} = \Gamma_t(\mathbf a_{ij}^{(\mathrm e)}),
\end{equation}
where $\sigma$ is the sigmoid function, $t$ is the current epoch, $T$ is the total number of epochs, $\gamma$ controls the final temperature, $\Gamma_t(\cdot)$ acts element-wise on vectors, and $\mathbf M_{ij:}$ is row $j$ of $\mathbf M_i$.
Subsequently, the tensor is configured using Eq.\eqref{eq:configured-embedding-problem} and passed to the sample selector:
\begin{equation}
\label{eq:sample-configuration-gates}
        s_i = \sigma(\Psi(\operatorname{vec}(\operatorname{sg}(\widetilde{\mathbf H}_i)))),
\end{equation}
where $\operatorname{vec}(\cdot)$ vectorizes its matrix argument. $\Psi(\cdot):\mathbb R^{Fd}\rightarrow\mathbb R$ is a lightweight multilayer perceptron. 
Eq.\eqref{eq:embedding-configuration-gates} turns the two logits into decisions in the order they affect learning.
The scalar $g_{ij}$ gates the whole feature embedding, whereas the $k$-th entry $M_{ijk}$ gates one unit within it.
Together they form $\widetilde{\mathbf H}_i$. 
$\Psi(\cdot)$ then evaluates this already-configured tensor to produce $s_i$. 
Thus, the feature and embedding-unit selectors decide what the backbone observes, and the sample selector decides how strongly that resulting example contributes to the loss.

The search step $t\in[0, T]$ progresses over a horizon $T$, and $\gamma>1$ gradually sharpens the feature and embedding-unit gates. 
During the early search stage, continuous values enable gradient propagation to competing configurations, whereas the logits gradually become hard decisions in the later stage for retraining.
Eqs.\eqref{eq:selector-logits}--\eqref{eq:sample-configuration-gates} therefore instantiate $s_i,\mathbf g_i,\mathbf M_i$ in the unified formulation. 
Note that all tables, unit representations, and sample scorers are shared across the dataset rather than learned independently for each sample.

\subsection{Cross-Level Coordination via OT}
\label{sec:selection-weighted-ot}
The common information path couples the selectors indirectly through prediction, but it still does not say which fine-grained decisions support a coarse one. 
A sample can receive a large weight while placing most of its feature mass on poorly matched fields; an active feature can likewise retain incompatible units.
However, comparing masks entry by entry is not meaningful here because a sample score $s_i$, a feature weight vector $\mathbf{g}_i$, and a unit weight matrix $\mathbf{M}_i$ live on different supports.
We instead read each relation as a coarse-to-fine transport. A coarse selection score $\alpha$ determines the mass to be explained, and a normalized fine-level allocation $\boldsymbol\pi$ distributes that mass over representations $\{\mathbf z_v\}$. 
For ground cost $c$, the transport between the corresponding finite measures has the closed form, for $\alpha\geq0$, $\pi_v\geq0$, and $\sum_v\pi_v=1$, formalized as:
\begin{equation}
\label{eq:closed-form-ot}
    \operatorname{OT}_{c}\big(
        \alpha\delta_{\mathbf z},
        \alpha\sum_v\pi_v\delta_{\mathbf z_v}
   \big)
    =
    \alpha\sum_v\pi_v c(\mathbf z,\mathbf z_v),
\end{equation}
where $\delta_{\mathbf z}$ is a Dirac measure. 
For given selection outputs, the single coarse support and fixed destination marginal uniquely determine $T_v^\star=\alpha\pi_v$.
The exact OT cost in Eq.~\eqref{eq:closed-form-ot}~\cite{choi2025overcoming, ge2021ota} thus acts as a selection-weighted representation-compatibility penalty, with mass, allocation, and representations jointly learned during search.

For example, in the sample--feature relation, $s_i$ supplies the coarse mass, the normalized feature gates distribute it, and the sample and feature representations define the transport cost. 
The other two relations use the same pattern at their own granularity.
Fine-level scores determine a unit-mass destination profile, while the coarse score $\alpha$ alone determines how much mass is transported.
For a positive vector $\mathbf q$, we use the temperature-normalized selection weights, expressed as follows:
\begin{equation}
    \left[\mathcal N_{\tau}(\mathbf q)\right]_r
    =
    \frac{
        \exp\!\left(\log(q_r+\varepsilon)/\tau\right)
    }{
        \sum_l
        \exp\!\left(\log(q_l+\varepsilon)/\tau\right)
    },
\end{equation}
\begin{equation}
\label{eq:transport-allocations}
    \boldsymbol\pi_i^{(\mathrm{sf})} =\mathcal N_{\tau}(\mathbf g_i), \quad
    \boldsymbol\pi_{ij}^{(\mathrm{fe})} =\mathcal N_{\tau}(\mathbf M_{ij:}), \quad
    \boldsymbol\pi_i^{(\mathrm{se})} =\mathcal N_{\tau} (\operatorname{vec}(\mathbf M_i)),
\end{equation}
where tiny $\varepsilon >0$ prevents $\log 0$, and $\tau>0$ controls concentration. 
We evaluate the normalization with log-sum-exp. In the limit $\varepsilon \rightarrow0$, a gate that vanishes relative to non-vanishing alternatives receives vanishing mass instead of being treated as an ordinary bounded logit.
The indices $r$ and $l$ range over entries of $\mathbf q$. We index the last allocation as $\pi_{ijk}^{(\mathrm{se})}$ after reshaping it to $F\times d$.

Then, semantic compatibility is measured in a shared representation space of dimension $d$. 
We use the pooled sample representation and cosine ground cost, formulated as follows:
\begin{equation}
    \overline{\mathbf h}_i
    =
    \frac{1}{F}\sum_{j=1}^{F}\mathbf h_{ij},
    \qquad
    c(\mathbf a,\mathbf b)
    =
    1-
    \frac{\mathbf a^\top\mathbf b}
    {(\lVert\mathbf a\rVert_2+\varepsilon)
     (\lVert\mathbf b\rVert_2+\varepsilon)},
    \label{eq:transport-ground-cost}
\end{equation}
where tiny $\varepsilon > 0$ also keeps the cosine cost defined for near-zero representations, and $\|\cdot\|_2$ is the $l_2$ norm.
Recall that $\mathbf h_{ij}$ represents feature $j$ and $\mathbf u_{jk}$, the $k$-th column of $\mathbf U_j$, represents its $k$-th embedding unit. We use the unmasked embeddings here as stable semantic anchors while the configuration changes. 
The stop-gradient in Eq.\eqref{eq:selector-logits} applies only to the unit-logit scoring path, while the representation costs remain differentiable.
Instantiating Eq.\eqref{eq:closed-form-ot} gives the three costs needed as:
\begin{equation}
    \begin{aligned}
        \mathcal W_{\mathrm{sf}}(i)
        &=
        s_i\sum_{j=1}^{F}
        \pi_{ij}^{(\mathrm{sf})}
        c(\overline{\mathbf h}_i,\mathbf h_{ij}),\\
        \mathcal W_{\mathrm{fe}}(i)
        &=
        \frac{1}{F}\sum_{j=1}^{F}g_{ij}
        \sum_{k=1}^{d}
        \pi_{ijk}^{(\mathrm{fe})}
        c(\mathbf h_{ij},\mathbf u_{jk}),\\
        \mathcal W_{\mathrm{se}}(i)
        &=
        s_i\sum_{j=1}^{F}\sum_{k=1}^{d}
        \pi_{ijk}^{(\mathrm{se})}
        c(\overline{\mathbf h}_i,\mathbf u_{jk}).
    \end{aligned}
    \label{eq:three-transport-costs}
\end{equation}
The first line transports sample mass to features, the second transports each feature's mass to its units, and the third provides a direct sample-to-unit path. Substituting these terms into Eq.\eqref{eq:coordination-target} gives an explicit signal to all three relations. 
Each term penalizes a selected coarse component when its selector-induced fine allocation has high representation cost.
This objective coordinates relative semantic allocation, while the CTR prediction loss and the selection constraints determine the absolute gate magnitudes.

\subsection{Optimization}
\label{sec:optimization}
Next, we will introduce how OptIDC is optimized, which can be divided into two stages: search and retrain.

\noindent\textbf{Search Stage.}
Configuration search must avoid two trivial outcomes: retaining every feature and unit, or suppressing nearly all samples. 
We therefore use sparsity on the fine-grained gates and control sample selection with a ratio constraint~\cite{SSIM}. 
Sparsity is applied directly to selector outputs rather than to their raw lookup parameters. 
In practice, optimization is performed using mini-batches: sample-indexed sums and averages are evaluated over the current mini-batch, while lookup-table regularizers are computed over the full tables. For clarity, we present the regularization structure using dataset-level notation:
\begin{equation}
\label{eq:selection-regularization}
    \begin{aligned}
        \Omega_{\mathrm{sp}}
        &=
        \lambda_{\mathrm f}
        \lVert\Gamma_t(\mathbf{A}^{(f)})\rVert_1
        +
       \lambda_{\mathrm e}\big(
        \sum_{i=1}^{N}\lVert\mathbf M_i\rVert_1 + \lVert\Gamma_t(\mathbf{B}^{(e)})\rVert_1\big),\\
        \Omega_{\mathrm s}
        &=
        \lambda_{\mathrm{rat}}
        \left(
            \frac{1}{N}
            \sum_{i=1}^{N}s_i-\rho
        \right)^2
        +
        \frac{\lambda_{\mathrm{ent}}}{N}
        \sum_{i=1}^{N}h_{\mathrm{bin}}(s_i),
    \end{aligned}
\end{equation} 
where $\lambda_{\mathrm f}$ and $\lambda_{\mathrm e}$ are sparsity regularization weights for feature and embedding-unit masks, $\lambda_{\mathrm{rat}}$ and $\lambda_{\mathrm{ent}}$ are ratio and entropy regularization weights for sample masks, $\rho$ is an expected sample-retention ratio (e.g., 0.98) and $h_{\mathrm{bin}}(z)=-z\log z-(1-z)\log(1-z)$ is binary entropy to encourage binarization of the sample masks. 
Its arguments are numerically clamped in implementation. 
Finally, combining Eqs.~\eqref{eq:unified-recommendation}, \eqref{eq:coordination-target}, and \eqref{eq:selection-regularization} gives the search stage objective:
\begin{equation}
    \min_{\theta,\Theta,\boldsymbol\varphi}
    \mathcal L_{\mathrm{search}}
    =
    \mathcal L_{\mathrm{ctr}}
    +\mathcal L_{\mathrm{coord}}
    +\Omega_{\mathrm{sp}}
    +\Omega_{\mathrm s},
    \label{eq:search-objective}
\end{equation}
where $\Omega_{\mathrm{sp}}+\Omega_{\mathrm s}$ instantiates $\Omega_{\mathrm{sel}}$, and Eq.\eqref{eq:three-transport-costs} instantiates the three components of $\mathcal L_{\mathrm{coord}}$. 
Setting the coordination coefficient $\lambda_{\mathrm{co}}$ to zero recovers IDC. 
The selector parameters $\boldsymbol\varphi$ collect $\mathbf A^{(\mathrm f)}$, $\mathbf B^{(\mathrm e)}$, $\{\mathbf U_j\}_{j=1}^{F}$, and the parameters of $\Psi(\cdot)$.
Note that during training, parameters are actually updated using mini-batches of samples.

\noindent\textbf{Retrain Stage.}
Search identifies a configuration, and final fitting estimates the model on that fixed input space. 
Let $s_i^\star,\mathbf g_i^\star,\mathbf M_i^\star$ denote the selector outputs obtained by solving Eq.\eqref{eq:search-objective}.
Sample, feature, and unit gates use the natural midpoint of their sigmoid range:
\begin{equation}
\label{eq:hard-configuration}
\widehat s_i = \mathbb I(s_i^\star >\tfrac12), \quad
\widehat{\mathbf g}_i = \mathbb I(\mathbf g_i^\star > \tfrac12), \quad
\widehat{\mathbf M}_i = \mathbb I(\mathbf M_i^\star > \tfrac12),
\end{equation}
where $\mathbb{I}(\cdot)$ is the element-wise indicator function. 
Because the feature and unit scores are value-indexed, their hard decisions are materialized as lookup tables, while sample decisions define the retained training set $\widehat{\mathcal D}  = \{ (\mathbf{x}_i, y_i) \mid \widehat{s}_i = 1, (\mathbf{x}_i, y_i) \in \mathcal{D} \}$.
Then, we remove the auxiliary selectors and coordination loss, initialize the model parameters $(\Theta,\theta)$ from the checkpoint saved at epoch $T_c$ of the search stage, and retrain with all hard masks fixed:
\begin{equation}
\label{eq:retrain-objective}
\min_{\theta,\Theta} \mathcal L_{\mathrm{retrain}} = 
\frac{1}{|\widehat{\mathcal D}|}
        \sum_{i\in\widehat{\mathcal D}}
        \ell\!\left(
            y_i,
            f_{\theta}
            (\widehat{\mathbf H}_i)
        \right), \; 
        \widehat{\mathbf H}_i = \mathbf H_i \odot(\widehat{\mathbf g}_i\mathbf 1_d^\top) \odot\widehat{\mathbf M}_i.
\end{equation}
Therefore, search decides what information to keep, and retraining estimates the embedding and backbone parameters for that choice. 
At inference time, a new sample retrieves the feature and unit masks of its active values. The sample selector is used only to construct the retraining set and introduces no serving-time overhead.

%% file: sec/5_experiments.tex
\section{Experiments}\label{sec:experiments}
In this section, we conduct experiments to answer the following five key questions.
\begin{itemize}[leftmargin=*]
    \item RQ1: How does OptIDC perform compared to the baselines?
    \item RQ2: What do unified configuration and cross-level coordination each contribute?
    \item RQ3: How does the efficiency of OptIDC compare to the baseline?
    \item RQ4: How do the cost construction and coordination methods affect OptIDC?
    \item RQ5: How effectively does OptIDC mitigate selection conflicts?
\end{itemize}

\subsection{Experiment Setup}\label{sec:experiments:setup}
\noindent\textbf{Datasets.}
To comprehensively evaluate the effectiveness of OptIDC, we follow prior work~\cite{erase} and conduct experiments on four public datasets: Avazu\footnote{https://www.kaggle.com/c/avazu-ctr-prediction}, Criteo\footnote{https://www.kaggle.com/c/criteo-display-ad-challenge}, MovieLens-1M\footnote{https://grouplens.org/datasets/movielens/}, and AliCCP\footnote{https://tianchi.aliyun.com/dataset/408}.
Avazu and Criteo are widely used in CTR for studying feature and dimension selection~\cite{optfs, PEP}. 
MovieLens-1M is a representative benchmark dataset commonly adopted to validate different selection methods. 
In addition, to assess performance in settings closer to real-world prediction systems, we include the AliCCP dataset, which is collected from the Taobao e-commerce platform and is widely used for CTR estimation.
Detailed preprocessing procedures and dataset statistics are provided in Appendix~\ref{sec:preprocessing}.
\begin{table*}[htbp]
\centering
\caption{Results on Avazu and Criteo (mean $\pm$ standard deviation over five runs with different random seeds), where the best and second best results are marked in bold and underlined, respectively. Note that $^{*}$ indicates a significance level of $p\leq 0.05$ based on a two-sample t-test between our method and the best baseline.}
\vspace{-7pt}
\setlength{\tabcolsep}{1pt}
\resizebox{1.0\textwidth}{!}{
\begin{tabular}{c|c|cccccccccc}
\specialrule{0.1em}{1pt}{1pt}
    \multirow{2}{*}{Dataset}
        &  \multirow{2}{*}{Model}
            & \multicolumn{2}{c|}{DNN} & \multicolumn{2}{c|}{DeepFM} & \multicolumn{2}{c|}{DCN} & \multicolumn{2}{c|}{AutoInt} & \multicolumn{2}{c}{FinalMLP} \\
    \cline{3-12}
    & & AUC$\uparrow$ & LogLoss$\downarrow$ & AUC$\uparrow$ & LogLoss$\downarrow$ & AUC$\uparrow$ & LogLoss$\downarrow$ & AUC$\uparrow$ & LogLoss$\downarrow$ & AUC$\uparrow$ & LogLoss$\downarrow$ \\
   \midrule
\multirow{12}{*}{\rotatebox{90}{Avazu}}
& Backbone
& .7893$\pm$.0001 & .3757$\pm$.0001
& .7896$\pm$.0004 & .3759$\pm$.0004
& .7893$\pm$.0004 & .3759$\pm$.0004
& .7895$\pm$.0003 & .3755$\pm$.0003
& .7898$\pm$.0004 & .3740$\pm$.0003 \\
& AutoField
& .7840$\pm$.0007 & .3812$\pm$.0007
& .7842$\pm$.0003 & .3809$\pm$.0008
& .7838$\pm$.0006 & .3806$\pm$.0004
& .7838$\pm$.0002 & .3815$\pm$.0005
& .7835$\pm$.0006 & .3812$\pm$.0015 \\
& LPFS
& .7888$\pm$.0007 & .3762$\pm$.0009
& .7896$\pm$.0003 & .3758$\pm$.0003
& .7894$\pm$.0003 & .3758$\pm$.0003
& .7894$\pm$.0004 & .3756$\pm$.0004
& .7898$\pm$.0002 & .3739$\pm$.0001 \\
& OptEmbed
& .7884$\pm$.0008 & .3766$\pm$.0009
& .7895$\pm$.0004 & .3756$\pm$.0003
& .7885$\pm$.0007 & .3765$\pm$.0007
& .7884$\pm$.0009 & .3768$\pm$.0010
& .7885$\pm$.0007 & .3757$\pm$.0011 \\
& OptFS
& .7907$\pm$.0002 & .3736$\pm$.0001
& .7915$\pm$.0002 & .3731$\pm$.0001
& .7908$\pm$.0001 & .3735$\pm$.0001
& .7904$\pm$.0003 & .3738$\pm$.0001
& .7916$\pm$.0005 & .3730$\pm$.0003 \\
& PEP
& .7892$\pm$.0005 & .3754$\pm$.0008
& .7895$\pm$.0005 & .3755$\pm$.0006
& .7891$\pm$.0001 & .3745$\pm$.0001
& .7892$\pm$.0006 & .3755$\pm$.0006
& .7894$\pm$.0002 & .3743$\pm$.0003 \\
& AutoDim
& .7898$\pm$.0005 & .3744$\pm$.0005
& .7899$\pm$.0006 & .3744$\pm$.0005
& .7892$\pm$.0006 & .3749$\pm$.0003
& .7899$\pm$.0005 & .3746$\pm$.0006
& .7909$\pm$.0005 & .3736$\pm$.0004 \\
& ES
& .7908$\pm$.0001 & .3735$\pm$.0001
& .7915$\pm$.0002 & .3731$\pm$.0001
& .7909$\pm$.0002 & \underline{.3734}$\pm$.0001
& .7907$\pm$.0001 & .3736$\pm$.0001
& .7915$\pm$.0002 & .3730$\pm$.0001 \\
& AutoDenoise
& .7834$\pm$.0002 & .3820$\pm$.0002
& .7835$\pm$.0003 & .3819$\pm$.0002
& .7832$\pm$.0001 & .3819$\pm$.0004
& .7831$\pm$.0004 & .3820$\pm$.0005
& .7836$\pm$.0003 & .3820$\pm$.0005 \\
& DS
& .7899$\pm$.0004 & .3745$\pm$.0006
& .7907$\pm$.0003 & .3736$\pm$.0006
& .7899$\pm$.0006 & .3744$\pm$.0005
& .7898$\pm$.0002 & .3745$\pm$.0003
& .7919$\pm$.0002 & .3731$\pm$.0008 \\
& IDC
& \underline{.7915}$\pm$.0003 & \underline{.3732}$\pm$.0003
& \underline{.7918}$\pm$.0003 & \underline{.3729}$\pm$.0001
& \underline{.7914}$\pm$.0002 & \underline{.3734}$\pm$.0003
& \underline{.7915}$\pm$.0002 & \underline{.3731}$\pm$.0003
& \underline{.7925}$\pm$.0001 & \underline{.3728}$\pm$.0002 \\
& OptIDC
& \textbf{.7930}$^{*}$$\pm$.0001 & \textbf{.3722}$^{*}$$\pm$.0001
& \textbf{.7933}$^{*}$$\pm$.0001 & \textbf{.3721}$^{*}$$\pm$.0001
& \textbf{.7931}$^{*}$$\pm$.0002 & \textbf{.3721}$^{*}$$\pm$.0002
& \textbf{.7934}$^{*}$$\pm$.0002 & \textbf{.3721}$^{*}$$\pm$.0001
& \textbf{.7941}$^{*}$$\pm$.0002 & \textbf{.3719}$^{*}$$\pm$.0003 \\

\midrule
\multirow{12}{*}{\rotatebox{90}{Criteo}}
& Backbone
& .8097$\pm$.0001 & .4419$\pm$.0001
& .8088$\pm$.0003 & .4435$\pm$.0001
& .8094$\pm$.0003 & .4425$\pm$.0001
& .8093$\pm$.0002 & .4424$\pm$.0002
& .8097$\pm$.0003 & .4420$\pm$.0002 \\
& AutoField
& .8085$\pm$.0004 & .4433$\pm$.0004
& .8083$\pm$.0004 & .4434$\pm$.0004
& .8084$\pm$.0005 & .4433$\pm$.0005
& .8086$\pm$.0002 & .4432$\pm$.0002
& .8091$\pm$.0003 & .4425$\pm$.0004 \\
& LPFS
& .8107$\pm$.0001 & .4410$\pm$.0001
& .8104$\pm$.0001 & .4413$\pm$.0001
& .8098$\pm$.0002 & .4418$\pm$.0002
& .8097$\pm$.0002 & .4418$\pm$.0003
& .8102$\pm$.0001 & .4414$\pm$.0001 \\
& OptEmbed
& .8105$\pm$.0001 & .4412$\pm$.0002
& .8108$\pm$.0001 & .4410$\pm$.0001
& .8104$\pm$.0001 & .4413$\pm$.0001
& .8104$\pm$.0002 & .4413$\pm$.0002
& .8099$\pm$.0001 & .4415$\pm$.0002 \\
& OptFS
& .8107$\pm$.0001 & .4410$\pm$.0001
& .8105$\pm$.0001 & .4411$\pm$.0001
& .8108$\pm$.0001 & .4409$\pm$.0001
& .8107$\pm$.0001 & .4410$\pm$.0001
& .8108$\pm$.0001 & .4411$\pm$.0001 \\
& PEP
& .8105$\pm$.0001 & .4410$\pm$.0001
& .8105$\pm$.0001 & .4411$\pm$.0001
& .8106$\pm$.0001 & .4410$\pm$.0001
& .8105$\pm$.0000 & .4410$\pm$.0001
& .8091$\pm$.0005 & .4427$\pm$.0007 \\
& AutoDim
& .8085$\pm$.0009 & .4429$\pm$.0009
& .8083$\pm$.0008 & .4431$\pm$.0008
& .8078$\pm$.0012 & .4436$\pm$.0012
& .8086$\pm$.0010 & .4431$\pm$.0011
& .8084$\pm$.0011 & .4431$\pm$.0009 \\
& ES
& .8107$\pm$.0001 & .4410$\pm$.0001
& .8105$\pm$.0002 & .4411$\pm$.0002
& .8108$\pm$.0001 & .4409$\pm$.0001
& .8106$\pm$.0001 & .4411$\pm$.0002
& .8107$\pm$.0001 & .4411$\pm$.0001 \\
& AutoDenoise
& .8104$\pm$.0002 & .4415$\pm$.0002
& .8104$\pm$.0002 & .4417$\pm$.0001
& .8103$\pm$.0002 & .4418$\pm$.0003
& .8101$\pm$.0001 & .4420$\pm$.0001
& .8104$\pm$.0002 & .4415$\pm$.0001 \\
& DS
& .8107$\pm$.0001 & .4410$\pm$.0001
& .8104$\pm$.0001 & .4413$\pm$.0001
& .8106$\pm$.0001 & .4412$\pm$.0001
& .8104$\pm$.0001 & .4412$\pm$.0001
& .8106$\pm$.0000 & .4411$\pm$.0001 \\
& IDC
& \underline{.8110}$\pm$.0001 & \underline{.4408}$\pm$.0001
& \underline{.8110}$\pm$.0001 & \underline{.4407}$\pm$.0001
& \underline{.8110}$\pm$.0001 & \underline{.4408}$\pm$.0003
& \underline{.8110}$\pm$.0001 & \underline{.4407}$\pm$.0001
& \underline{.8110}$\pm$.0001 & \underline{.4407}$\pm$.0002 \\
& OptIDC
& \textbf{.8113}$^{*}$$\pm$.0001 & \textbf{.4404}$^{*}$$\pm$.0001
& \textbf{.8114}$^{*}$$\pm$.0001 & \textbf{.4404}$^{*}$$\pm$.0001
& \textbf{.8114}$^{*}$$\pm$.0001 & \textbf{.4404}$^{*}$$\pm$.0001
& \textbf{.8113}$^{*}$$\pm$.0001 & \textbf{.4404}$^{*}$$\pm$.0001
& \textbf{.8115}$^{*}$$\pm$.0001 & \textbf{.4402}$^{*}$$\pm$.0001 \\
    \specialrule{0.1em}{1pt}{1pt}
\end{tabular}}
\label{tab:main_result}
\end{table*}
\noindent\textbf{Metrics.}
Following prior work~\cite{optfs}, we adopt two standard evaluation metrics for CTR prediction: area under the ROC curve (AUC) and cross-entropy loss (LogLoss). Small AUC improvements can be practically meaningful in CTR prediction~\cite{DeepFM, optfs, yin2025feature}. 

\noindent\textbf{Baseline Methods and Backbone Methods.}
We select representative methods from prior studies on input-side information selection methods in deep CTR prediction. 
Specifically, we compare OptIDC with feature selection baselines including AutoField~\cite{autofield}, LPFS~\cite{LPFS}, and OptFS~\cite{optfs}; embedding configuration baselines AutoDim~\cite{AutoDim}, OptEmbed~\cite{optembed}, and PEP~\cite{PEP}; and the sample selection baseline AutoDenoise~\cite{Autodenoise}. 
In addition, we compare OptIDC with single-level variants of our configuration path: feature selection (FS), embedding-unit selection (ES), and sample selection (DS). We also report IDC, which activates all three levels but removes cross-level coordination.
Note that the feature selection (FS) in OptIDC is identical to that in OptFS.
To evaluate generalization, each method is combined with five main backbone models: DNN, DeepFM~\cite{DeepFM}, DCN~\cite{wang2017deep}, AutoInt~\cite{song2019autoint}, and FinalMLP~\cite{finalmlp}. We also include a base model without any selection method, denoted as `Backbone'. 
Implementation details are provided in Appendix~\ref{sec:implementation}.

\subsection{RQ1: Performance Comparison}\label{sec:experiments:rq1}
This subsection evaluates OptIDC against multiple baseline methods spanning three types of input-side information selection methods: sample selection, feature selection, and embedding configuration.
In Table~\ref{tab:main_result}, we report the overall performance of OptIDC and other baseline methods across two benchmark datasets, evaluated with five different backbone models. The results of the other two datasets are shown in Table~\ref{tab:appendix:main_result} in Appendix~\ref{sec:appendix:performance}. Based on the experimental results, we make the following observations:
1) Several strong single-level methods (e.g., OptFS, LPFS, ES, and DS) generally outperform the Backbone, confirming that input-side configuration is useful in its own right. 
The improvement is not uniform, however: a method that works well for one backbone or dataset may offer little benefit for another. This variation is
consistent with our premise that samples, feature values, and embedding units expose different kinds of redundancy.
2) IDC is generally stronger than its single-level counterparts, showing that the three decisions provide complementary information even when they are connected only by the prediction objective. 
3) Our proposed OptIDC achieves significant AUC improvements over IDC on all datasets, with a clearer margin on Avazu. This result highlights the importance of introducing the cross-level coordination mechanism, thereby learning a more effective input-side configuration.

\subsection{RQ2: Ablation Study}\label{sec:experiments:rq2}
Next, we analyze the contributions of joint configuration and OT coordination separately.
Specifically, we remove different configuration paths or coordination terms from OptIDC, resulting in ten ablation variants.
These include joint sample and feature selection (`DSFS'), joint sample and embedding-unit selection (`DSES'), and joint feature and embedding-unit selection (`FSES'), together with their counterparts using the corresponding OT relation (`DSFS+OT', `DSES+OT', and `FSES+OT'); joint three-level selection without coordination (`IDC'); and three variants that individually remove $\mathcal W_{\mathrm{sf}}$, $\mathcal W_{\mathrm{se}}$, or $\mathcal W_{\mathrm{fe}}$ while keeping all three selectors active.
The single-level results for FS, ES, and DS are reported in Table~\ref{tab:main_result}, where FS corresponds to OptFS.
We additionally include AOP, which jointly trains AutoDenoise, OptFS, and PEP for three-level selection while preserving their key network modules. AOP's regularization weights are tuned on the validation set under joint training, while other hyperparameters retain their standalone settings.
The corresponding results are summarized in Table~\ref{tab:ablation_result}, and additional results on the Criteo dataset are shown in Table~\ref{tab:appendix:ablation_result} in Appendix~\ref{sec:appendix:ablation}.
Combining two levels generally improves upon the corresponding single-level variants, while IDC provides further gains in some settings. Adding the corresponding OT relation improves AUC and improves or maintains LogLoss for each two-level model across the five backbones.
OptIDC further improves both metrics over IDC, supporting the benefits of explicit coordination within joint three-level configuration.
Removing any individual OT term from OptIDC lowers AUC and increases LogLoss across all five backbones on Avazu. These results support the contribution of each coordination relation.
OptIDC also outperforms AOP, suggesting that its benefits extend beyond directly combining existing selection methods.

\begin{table}[htbp]
\centering
\caption{Training and inference time on the Avazu dataset.}
\vspace{-5pt}
\setlength{\tabcolsep}{2pt}
\resizebox{1.0\linewidth}{!}{
\begin{tabular}{c|cccccc}
\specialrule{0.1em}{1pt}{1pt}
        Times (sec) & Backbone & OptEmbed & OptFS & AutoDim & AutoDenoise & OptIDC  \\
    \hline
        Train (Total) & 1069 & 3378 & 1301 & 2613 & 2481 & 2575  \\
        Train (Batch) & 0.0847 & 0.0930 & 0.0902 & 0.1880 & 0.0989 & 0.1032  \\
        Inference & 0.0034 & 0.0042 & 0.0040 & 0.0124 & 0.0040 & 0.0041 \\
\specialrule{0.1em}{1pt}{1pt}
\end{tabular}}
\label{tab:times}
\vspace{-15pt}
\end{table}
\begin{table*}[htb]
\centering
\caption{Ablation analysis of our method on the Avazu dataset, with the best results marked in bold.}
\resizebox{1.0\textwidth}{!}{
\begin{tabular}{c|c|cccccccccc}
\specialrule{0.1em}{1pt}{1pt}
    \multirow{2}{*}{Dataset}
        &  \multirow{2}{*}{Method}
            & \multicolumn{2}{c|}{DNN} & \multicolumn{2}{c|}{DeepFM} & \multicolumn{2}{c|}{DCN} & \multicolumn{2}{c|}{AutoInt} & \multicolumn{2}{c}{FinalMLP} \\
    \cline{3-12}
    & & AUC$\uparrow$ & LogLoss$\downarrow$ & AUC$\uparrow$ & LogLoss$\downarrow$ & AUC$\uparrow$ & LogLoss$\downarrow$ & AUC$\uparrow$ & LogLoss$\downarrow$ & AUC$\uparrow$ & LogLoss$\downarrow$ \\
    \midrule
    \multirow{12}{*}{\rotatebox{90}{Avazu}}
    & AOP & 0.7912 & 0.3731 & 0.7916 & 0.3731 & 0.7911 & 0.3735 & 0.7911 & 0.3734 & 0.7920 & 0.3727 \\
    & DSFS & 0.7911 & 0.3736 & 0.7917 & 0.3732 & 0.7914 & 0.3734 & 0.7911 & 0.3737 & 0.7925 & 0.3728 \\
    & DSFS+OT & 0.7915 & 0.3733 & 0.7919 & 0.3731 & 0.7921 & 0.3728 & 0.7916 & 0.3734 & 0.7926 & 0.3725 \\
    & DSES & 0.7913 & 0.3738 & 0.7920 & 0.3729 & 0.7912 & 0.3734 & 0.7913 & 0.3732 & 0.7920 & 0.3731 \\
    & DSES+OT & 0.7915 & 0.3738 & 0.7921 & 0.3729 & 0.7917 & 0.3732 & 0.7923 & 0.3727 & 0.7921 & 0.3729 \\
    & FSES & 0.7910 & 0.3733 & 0.7921 & 0.3727 & 0.7912 & 0.3732 & 0.7913 & 0.3733 & 0.7922 & 0.3726 \\
    & FSES+OT & 0.7920 & 0.3727 & 0.7928 & 0.3722 & 0.7920 & 0.3727 & 0.7928 & \textbf{0.3721} & 0.7938 & 0.3720 \\
    & IDC & 0.7915 & 0.3732 & 0.7918 & 0.3729 & 0.7914 & 0.3734 & 0.7915 & 0.3731 & 0.7925 & 0.3728 \\
    & w/o $\mathcal W_{\mathrm{sf}}$ & 0.7924 & 0.3727 & 0.7928 & 0.3726 & 0.7924 & 0.3729 & 0.7927 & 0.3728 & 0.7932 & 0.3723 \\
    & w/o $\mathcal W_{\mathrm{se}}$ & 0.7929 & 0.3723 & 0.7932 & 0.3722 & 0.7929 & 0.3726 & 0.7930 & 0.3725 & 0.7936 & 0.3722 \\
    & w/o $\mathcal W_{\mathrm{fe}}$ & 0.7926 & 0.3727 & 0.7930 & 0.3724 & 0.7925 & 0.3730 & 0.7927 & 0.3723 & 0.7935 & 0.3722 \\
    & OptIDC & \textbf{0.7930} & \textbf{0.3722} & \textbf{0.7933} & \textbf{0.3721} & \textbf{0.7931} & \textbf{0.3721} & \textbf{0.7934} & \textbf{0.3721} & \textbf{0.7941} & \textbf{0.3719} \\
    \specialrule{0.1em}{1pt}{1pt}
\end{tabular}}
\label{tab:ablation_result}
\end{table*}

\subsection{RQ3: Efficiency Analysis}\label{sec:experiments:rq3}
\begin{figure}[htbp]
    \centering
    \includegraphics[width=0.48\textwidth]{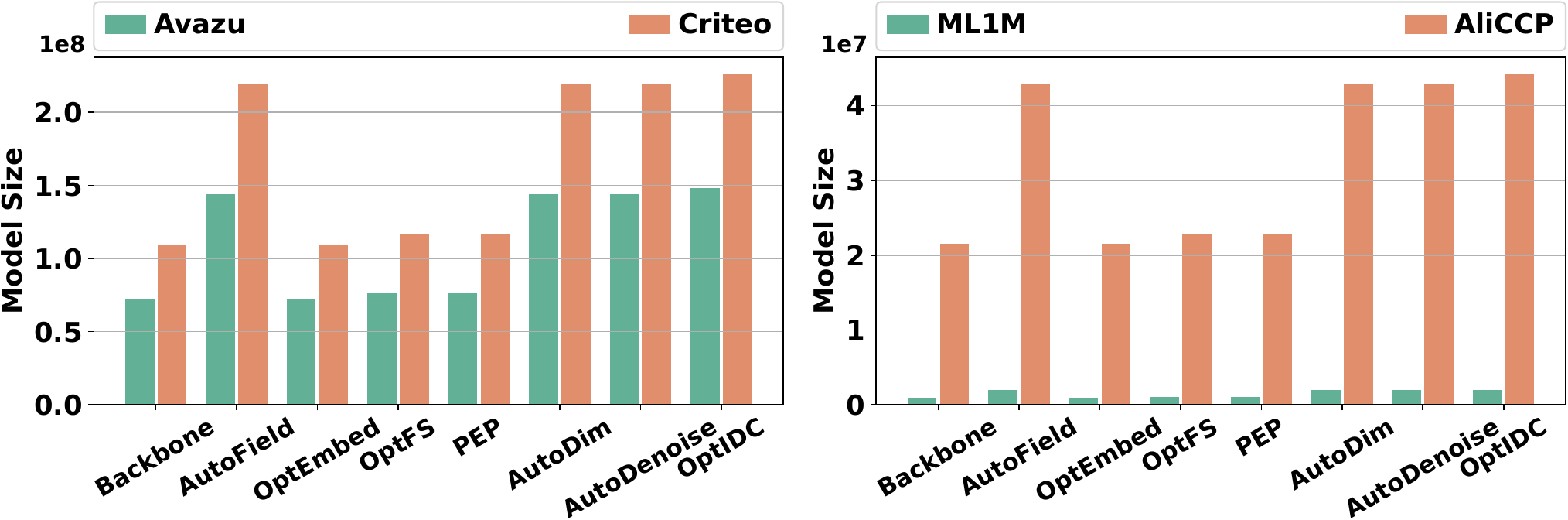}
    \Description[<Figure 4. Fully described in the text.>]{<A full description of Figure 4 can be found in Section 5.4.>}
    \caption{Model size of OptIDC and baselines on four datasets.}
    \label{fig:model_size}
    \vspace{-10pt}
\end{figure}

In this subsection, we evaluate the efficiency of OptIDC from the perspectives of model size and computational cost, using DNN as the backbone model.
As shown in Fig.~\ref{fig:model_size}, OptIDC has more parameters than Backbone, OptFS, PEP, and OptEmbed, while remaining comparable to AutoField, AutoDim, and AutoDenoise. Notably, unlike methods that focus on a single-level selection, OptIDC jointly models the three-level selection with a well-controlled parameter overhead and achieves improved performance.
In terms of efficiency, Table~\ref{tab:times} reports total training time and per-batch training and inference time on the Avazu dataset.
Total training time includes search and retraining where applicable.
OptIDC requires 2575 seconds in total, less than OptEmbed and AutoDim but more than Backbone, OptFS, and AutoDenoise.
Its per-batch training time is 0.1032 seconds, which is 4.3\%--14.4\% higher than OptEmbed, OptFS, and AutoDenoise, but lower than AutoDim.
Its inference time is 0.0041 seconds per batch, comparable to these three selection methods, versus 0.0034 seconds for the Backbone.
These results show that OptIDC supports joint three-level selection with moderate additional per-batch training cost relative to these single- and two-level selection methods.

\subsection{RQ4: In-depth Analysis of OptIDC}\label{subsec:indepth}
Next, we further analyze the role of OT coordination in OptIDC, focusing on the impact of different cost matrix constructions on OT coordination and the effectiveness of different coordination strategies.
Additionally,  selection ratio and hyperparameter sensitivity analyses are provided in Appendix~\ref{subsec:appendix:ratio} and Appendix~\ref{subsec:appendix:hyper}.

\begin{figure}[htbp]
  \centering
  \begin{subfigure}[t]{0.48\linewidth}
    \centering
    \includegraphics[width=\linewidth]{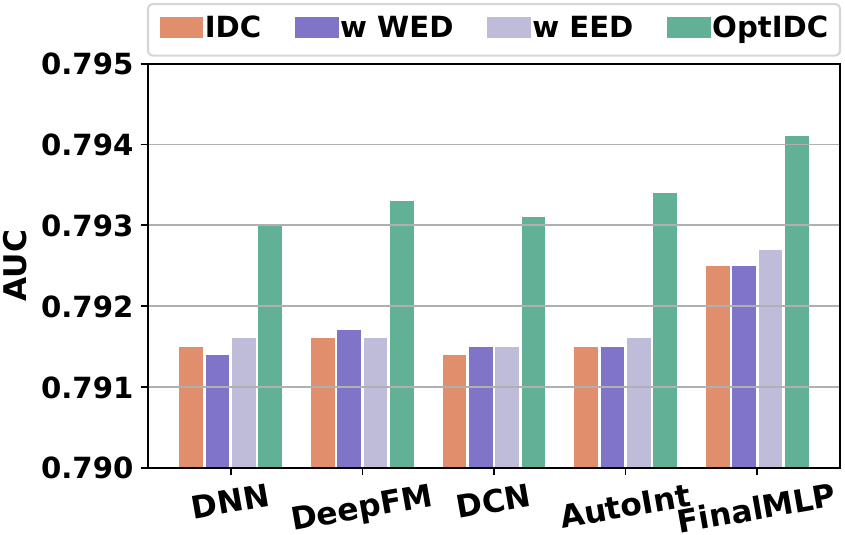}
    \caption{Impact of different cost matrix calculation methods.}
    \label{fig:OT_cost}
  \end{subfigure}
  \hspace{0.1em}
  \begin{subfigure}[t]{0.48\linewidth}
    \centering
\includegraphics[width=\linewidth]{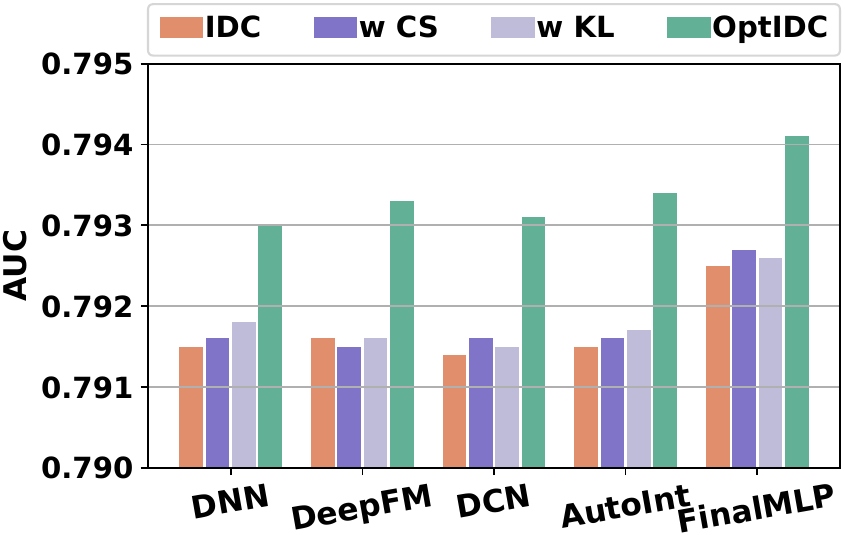}
    \caption{Effects of different coordination methods.}
    \label{fig:tls_am}
  \end{subfigure}
  \Description[<Figure 5. Fully described in the text.>]{<A full description of Figure 5 can be found in Section 5.5.>}
  \caption{Analysis of cost matrix design in OT coordination and different coordination methods on the Avazu dataset.}
  \label{fig:OT}
\end{figure}

\noindent\textbf{Impact of Different Cost Matrices.}
To examine how cost matrix design affects OT coordination, we compare several cost constructions on the Avazu dataset, including a variant without OT coordination (`IDC'), cost matrices based on weights Euclidean distance (`w WED') and embedding Euclidean distance (`w EED'), and the full OptIDC. 
Results in Fig.~\ref{fig:OT_cost} show that both Euclidean-distance-based costs yield limited gains, performing similarly to IDC. OptIDC achieves better performance, supporting the effectiveness of its cosine-based ground cost under the evaluated settings.

\noindent\textbf{Effects of different coordination methods.}
To investigate the impact of different coordination methods on model performance, we compare several coordination strategies on the Avazu dataset, including a variant without coordination (`IDC'), a cosine similarity-based coordination method (`w CS'), a symmetric KL divergence-based coordination method (`w KL'), and our proposed OptIDC. 
Definitions, normalization schemes, and tuning details of all alternatives are provided in
Appendix~\ref{sec:appendix:coordination-alternatives}.
The results are presented in Fig.~\ref{fig:tls_am}. 
As shown, both the cosine similarity-based and KL divergence-based coordination methods yield only marginal improvements, achieving performance comparable to that of IDC without any coordination methods. 
This suggests that directly matching selection weights provides limited benefits under the evaluated settings.
In contrast, OptIDC performs better across all five backbones, supporting the mass-preserving coarse-to-fine formulation used by our method. This comparison does not imply that OT discovers an unconstrained correspondence: the relation supports are fixed by the three configuration levels, and OT provides a common way to weight their compatibility.

\subsection{RQ5: Selection Consistency Analysis}\label{subsec:rq5}
\begin{figure}[htbp]
    \centering
    \includegraphics[width=0.48\textwidth]{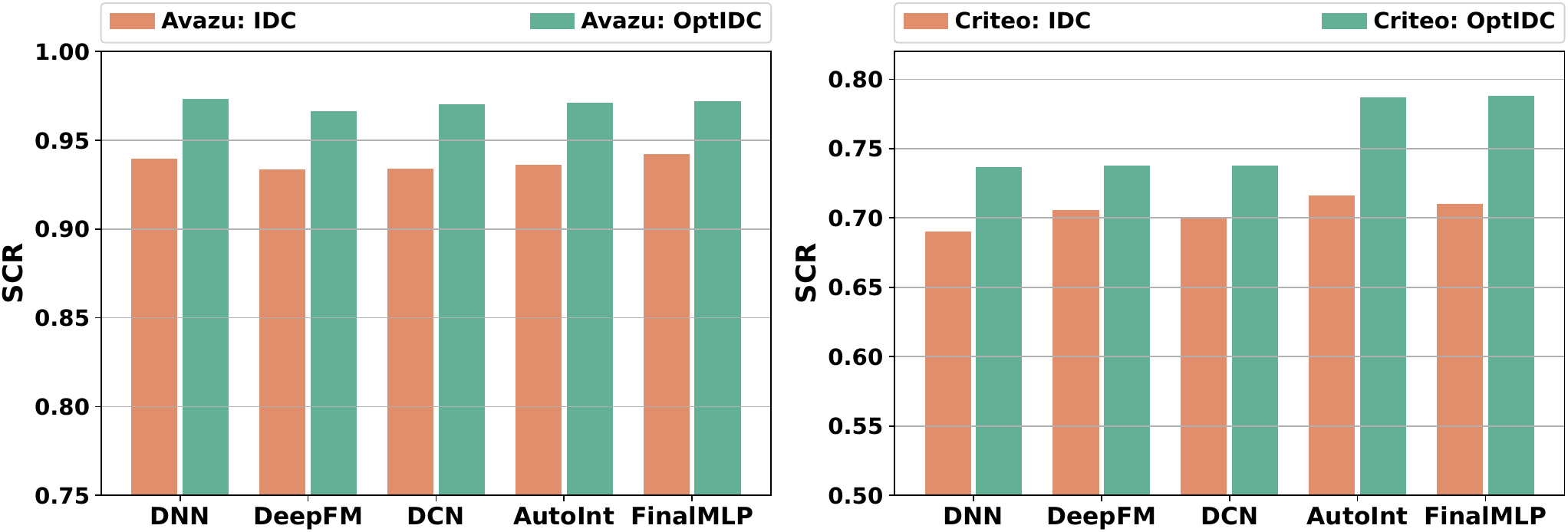}
    \Description[<Figure 6. Fully described in the text.>]{<A full description of Figure 6 can be found in Section 5.6.>}
    \caption{Selection consistency evaluation of OptIDC on the Avazu and Criteo datasets.}
    \label{fig:OT_F1}
\end{figure}

In this subsection, we evaluate OptIDC's ability to effectively mitigate selection conflicts and thereby improve selection consistency.
Due to the inherent differences in selection behaviors across different input-side levels, joint modeling can lead to selection conflicts, most directly reflected in discrepancies among the selection masks.
Here, we use SCR from Eq.~\eqref{eq:three-way-scr} as the selection consistency metric.
Since the selection masks at different levels operate at different granularities, they cannot be compared directly. Therefore, they are first mapped into a common space before the consistency is computed. The detailed mapping process is provided in Appendix~\ref{sec:mask mapping mechanism}.
As shown in Fig.~\ref{fig:OT_F1}, after introducing OT coordination, OptIDC has higher SCR than IDC for every backbone on both Avazu and Criteo. The result indicates greater agreement among the learned sample, feature, and embedding-unit configurations, which is consistent with the purpose of explicit cross-level coordination. We interpret it as supporting evidence rather than a causal explanation of the AUC gain, because SCR itself is not optimized by OptIDC.

%% file: sec/6_conclusion.tex
\section{Conclusions}\label{sec:conclusions}
In this paper, we investigate the input-side information configuration problem in deep CTR prediction and propose OptIDC, a unified framework for Optimal Input-side Dimensional Configuration.
OptIDC jointly models the selection mechanisms of samples, features, and embedding units, and introduces an Optimal Transport-based coordination strategy to coordinate information configuration across different levels.
Extensive experiments on four real-world datasets demonstrate that OptIDC can more effectively identify critical information and improve performance.

%% file: sec/7_appendix.tex
\section{Mask Mapping Mechanism}\label{sec:mask mapping mechanism}
In this appendix, $\mathbf G\in\{0,1\}^{n}$ and $\mathbf M\in\{0,1\}^{n\times d}$ denote the value-indexed hard feature and embedding-unit mask tables obtained after search, respectively, where $n=|\mathcal V|$.

\noindent\textbf{Feature frequency statistics.} 
Given the training set $\mathcal{D}$, for each feature $v_j$, $j \in \{1,2,\ldots, n\}$, we compute its occurrence frequency in the training set as:
\begin{equation}
\mathbf{F}_j^{\text{ori}} =  \sum_{i=1}^{|\mathcal{D}|} \mathbb{I}(v_j \in \mathbf{x}_i),
\end{equation}
where $\mathbb{I}(\cdot)$ is the indicator function, $\mathbf{F}^{\text{ori}}$ is the feature frequency vector for the original dataset $\mathcal{D}$. Moreover, to compute feature frequencies for selected samples, we first construct the selected sample set $\widehat{\mathcal{D}} = \{ (\mathbf{x}_i, y_i) \mid \widehat{{s}}_i = 1, (\mathbf{x}_i, y_i) \in \mathcal{D} \}$, where $\widehat{\mathbf{s}}$ is the sample mask and $|\widehat{\mathcal{D}}|$ is the number of selected samples.
Similarly, the frequency of the $j$-th feature in the selected sample set $\widehat{\mathcal{D}}$ is:
\begin{equation}
\mathbf{F}_j^{\text{sel}} = \sum_{i=1}^{|\widehat{\mathcal{D}}|} \mathbb{I}(v_j \in \mathbf{x}_i).
\end{equation}
\noindent\textbf{Sample Mask Mapping.} 
Sample selection itself does not directly operate on embeddings. Instead, it changes the empirical feature distribution of the retained samples. Since embeddings are updated only when their corresponding features appear, the feature occurrence frequency can be regarded as the activation frequency of embedding units. 
Thus, to quantify the impact of sample selection on embedding learning, we remap the sample selection mask $\mathbf{s}$ to an embedding-unit activation matrix $\mathbf{M}^\mathrm{s}$. This representation facilitates a direct comparison of information selection with feature selection and embedding-unit selection. Accordingly, $\mathbf{M}^\mathrm{s}$ can be expressed as follows:
\begin{equation}
\mathbf{M}^{\mathrm{s}} =   \log(\mathbf{F}^{\text{sel}} + 1) \mathbf{1}_d^\top, \quad \mathbf{M}^{\mathrm{s}} \in \mathbb{R}^{n \times d},
\end{equation}
where the $\log(x + 1)$ operation on $\mathbf{F}^{\text{sel}} \in \mathbb{R}^n$ is applied to reduce the influence of high-frequency features~\cite{wang2025universal}, which is also a commonly used practice in industrial CTR systems. And $\mathbf{1}_d \in \{1\}^d$ is an all-one vector of size $d$. This broadcasting operation projects the feature-level statistics into the embedding-unit space for unified comparison.

\noindent\textbf{Feature Mask Mapping.} 
Similarly, to further quantify the impact of feature selection on embedding learning, we remap the feature selection mask $\mathbf{G}$ to an embedding-unit activation matrix $\mathbf{M}^\mathrm{f}$, which can be denoted as:
\begin{equation}
\mathbf{w} = \log(\mathbf{F}^{\text{ori}} + 1),
\end{equation}

\begin{equation}
\mathbf{M}^{\mathrm{f}} = (\mathbf{w} \odot \mathbf{G}) \mathbf{1}_d^\top, \quad \mathbf{M}^{\mathrm{f}}  \in \mathbb{R}^{n \times d},
\end{equation}
where the $\log(x + 1)$ operation on $\mathbf{F}^{\text{ori}} \in \mathbb{R}^n$ is applied to reduce the influence of high-frequency features, $\odot$ is element-wise multiplication, and $\mathbf{1}_d \in \{1\}^d$ is an all-one vector of size $d$. 

\noindent\textbf{Embedding-unit Mask Mapping.} 
To quantify the impact of embedding-unit selection on embedding learning, we further consider the activation frequency of each embedding unit. Thus, the embedding-unit selection mask $\mathbf{M}$ is remapped to an embedding-unit activation matrix $\mathbf{M}^\mathrm{e}$, which can be formulated as follows:
\begin{equation}
\mathbf{M}^{\mathrm{e}} = (\mathbf{w} \mathbf{1}_d^\top) \odot \mathbf{M}, \quad \mathbf{M}^{\mathrm{e}} \in \mathbb{R}^{n \times d}.
\end{equation}

\section{Experimental Settings}
\subsection{Dataset Preprocessing and Statistics}
\label{sec:preprocessing}
For Avazu and Criteo, following prior work~\cite{optembed}, we filter low-frequency features using a threshold of 2. In Avazu, the instance ID field is removed, and the timestamp field is transformed into three fields: hour, weekday, and is\_weekend. Both datasets are then randomly split into training, validation, and test sets with an 8:1:1 ratio.
For MovieLens-1M, since the original dataset only contains ratings from 1 to 5, we convert it into a binary classification task by setting a rating threshold of 4, where $rating \geq 4$ indicates a click. As the dataset contains fewer non-ID features, no low-frequency filtering is applied, and the data are also randomly divided into training, validation, and test sets in an 8:1:1 ratio.
For AliCCP, following previous studies~\cite{MultiFS, SSIM}, we filter low-frequency features with a threshold of 10. We then randomly sampled 10\% of the original training set as the validation set and used the original test set for testing.
Table~\ref{tab:datasets} presents the statistics of the processed datasets.

\begin{table}
\centering
\caption{Dataset statistics.}
\resizebox{0.48\textwidth}{!}{
\begin{tabular}{c|c|c
|c|c}
\specialrule{0.1em}{1pt}{1pt}
        \multicolumn{1}{c|}{Dataset} & \multicolumn{1}{c|}{Avazu} & \multicolumn{1}{c|}{Criteo} & \multicolumn{1}{c|}{ML1M} & \multicolumn{1}{c}{AliCCP} \\
    \hline
        \#Train & 32,343,174 & 36,672,494 & 800,167 & 38,070,670  \\
        \#Validation & 4,042,897 & 4,584,061 & 100,021 & 4,229,235  \\
        \#Test & 4,042,896 & 4,584,062 & 100,021 & 43,016,614  \\
        \#Field & 24 & 39 & 7 & 18  \\
        \#Feature & 4,428,306 & 6,780,406 & 13,240 & 1,283,107  \\
\specialrule{0.1em}{1pt}{1pt}
\end{tabular}}
\label{tab:datasets}
\vspace{-10pt}
\end{table}

\subsection{Implementation Details}
\label{sec:implementation}
In this subsection, we describe the implementation details of OptIDC and all baseline methods. 
For shared hyperparameter settings, the embedding size and batch size are fixed to 16 and 4096, respectively.
The learning rate is selected from $\{$1e-3, 3e-4, 1e-4, 3e-5, 1e-5$\}$ and the $l_2$ regularization from $\{$3e-4, 1e-4, 3e-5, 1e-5, 3e-6, 1e-6$\}$.
For the MLP components of the backbone models, we adopt a three-layer fully connected architecture with hidden sizes $[1024, 512, 256]$.
All models are trained using the Adam optimizer with Xavier initialization.
For the hyperparameters of OptIDC, we select the optimal regularization penalty $\lambda_{\mathrm{rat}}$, $\lambda_{\mathrm{ent}}$, $\lambda_{\mathrm{f}}$, $\lambda_{\mathrm{e}}$ and final value $\gamma$ from $\{$5, 10, 50, 100, 500$\}$,$\{$0.01, 0.05, 0.1, 0.5, 1.0$\}$, $\{$1e-6, 1e-7, 1e-8, 1e-9, 1e-10$\}$, $\{$1e-7, 1e-8, 1e-9, 1e-10, 1e-11$\}$ and $\{$100, 200, 500, 1000$\}$, respectively, temperature coefficient $\tau$ is set to 0.1, and select the optimal OT penalty $\lambda_{\mathrm{co}}$ to $\{$1e-3, 1e-4, 1e-5, 1e-6, 1e-7$\}$. The expected sample-retention ratio is $\rho=0.98$.
During the retraining phase, the optimal learning rate and $l_2$ regularization can be re-selected, and the rewinding epoch $T_c$ is chosen from $\{$1,2,$\cdots$,$T-1$$\}$.
For baseline methods, we directly use the publicly available implementations of AutoField~\cite{autofield}, LPFS~\cite{LPFS}, OptFS~\cite{optfs}, PEP~\cite{PEP}, OptEmbed~\cite{optembed}, and AutoDenoise~\cite{Autodenoise}.
Since no official implementation of AutoDim~\cite{AutoDim} is available, we re-implement it following the specifications in the original paper.
For FinalMLP~\cite{finalmlp}, we follow its open-source implementation but disable its built-in feature selection module to isolate the effects of the input-side feature selection methods under comparison, while retaining its two-stream MLP architecture and interaction aggregation module. This backbone variant is used consistently for `Backbone', all baseline models, and OptIDC.
Note that all baselines are tuned for optimal parameter settings within the same hyperparameter range.
Each experiment is independently repeated with five different random seeds. Tables~\ref{tab:main_result} and~\ref{tab:appendix:main_result} report the mean $\pm$ standard deviation across the five runs, while the remaining experimental results are reported as means.

\section{More Experimental Results}\label{sec:appendix:analysis}

\subsection{Performance Comparison}\label{sec:appendix:performance}
This section evaluates OptIDC against multiple baseline methods spanning three types of input-side information selection methods: sample selection, feature selection, and embedding configuration.
In Table~\ref{tab:main_result} and Table~\ref{tab:appendix:main_result}, we report the overall performance of OptIDC and other baseline methods across four benchmark datasets, evaluated with five different backbone models.
Based on the experimental results, we make the following observations:
1) Several strong single-level methods (e.g., OptFS, LPFS, ES, and DS) generally outperform the Backbone, confirming that input-side configuration is useful in its own right. 
The improvement is not uniform, however: a method that works well for one backbone or dataset may offer little benefit for another. This variation is
consistent with our premise that samples, feature values, and embedding units expose different kinds of redundancy.
2) IDC is generally stronger than its single-level counterparts, showing that the three decisions provide complementary information even when they are connected only by the prediction objective. 
3) By contrast, our proposed OptIDC achieves significant AUC improvements over IDC on all datasets, with a clearer margin on Avazu. This result highlights the importance of introducing the cross-level coordination mechanism, thereby learning a more effective input-side configuration. On AliCCP, however, OptIDC achieves the highest AUC across all five backbones but does not achieve the lowest LogLoss, suggesting that its benefits are more consistently reflected in AUC.

\begin{table*}[htbp]
\centering
\caption{Results on ML1M and AliCCP (mean $\pm$ standard deviation over five runs with different random seeds), where the best and second best results are marked in bold and underlined, respectively. Note that $^{*}$ indicates a significance level of $p\leq 0.05$ based on a two-sample t-test between our method and the best baseline.}
\setlength{\tabcolsep}{1pt}
\resizebox{1.0\textwidth}{!}{
\begin{tabular}{c|c|cccccccccc}
\specialrule{0.1em}{1pt}{1pt}
    \multirow{2}{*}{Dataset}
        &  \multirow{2}{*}{Model}
            & \multicolumn{2}{c|}{DNN} & \multicolumn{2}{c|}{DeepFM} & \multicolumn{2}{c|}{DCN} & \multicolumn{2}{c|}{AutoInt} & \multicolumn{2}{c}{FinalMLP} \\
    \cline{3-12}
    & & AUC$\uparrow$ & LogLoss$\downarrow$ & AUC$\uparrow$ & LogLoss$\downarrow$ & AUC$\uparrow$ & LogLoss$\downarrow$ & AUC$\uparrow$ & LogLoss$\downarrow$ & AUC$\uparrow$ & LogLoss$\downarrow$ \\
    
    \midrule
    \multirow{12}{*}{\rotatebox{90}{ML1M}}
    & Backbone
    & .8151$\pm$.0001 & .5192$\pm$.0007
    & .8147$\pm$.0006 & .5201$\pm$.0015
    & .8148$\pm$.0004 & .5199$\pm$.0013
    & .8148$\pm$.0004 & .5205$\pm$.0020
    & .8138$\pm$.0003 & .5236$\pm$.0017 \\
    & AutoField
    & .8137$\pm$.0006 & .5211$\pm$.0010
    & .8137$\pm$.0003 & .5212$\pm$.0003
    & .8141$\pm$.0002 & .5206$\pm$.0012
    & .8137$\pm$.0003 & .5209$\pm$.0016
    & .8137$\pm$.0002 & .5225$\pm$.0011 \\
    & LPFS
    & .8154$\pm$.0006 & .5181$\pm$.0004
    & .8158$\pm$.0005 & .5194$\pm$.0014
    & .8157$\pm$.0004 & .5177$\pm$.0006
    & .8153$\pm$.0007 & .5184$\pm$.0010
    & .8136$\pm$.0008 & .5227$\pm$.0007 \\
    & OptEmbed
    & .8156$\pm$.0004 & .5188$\pm$.0007
    & .8155$\pm$.0006 & .5188$\pm$.0009
    & .8158$\pm$.0004 & .5176$\pm$.0007
    & .8159$\pm$.0004 & .5177$\pm$.0011
    & .8142$\pm$.0010 & .5209$\pm$.0012 \\
    & OptFS
    & .8161$\pm$.0002 & .5178$\pm$.0004
    & .8163$\pm$.0003 & .5173$\pm$.0004
    & .8164$\pm$.0001 & .5171$\pm$.0003
    & .8161$\pm$.0004 & .5173$\pm$.0007
    & .8156$\pm$.0003 & .5194$\pm$.0013 \\
    & PEP
    & .8068$\pm$.0012 & .5296$\pm$.0046
    & .8064$\pm$.0010 & .5287$\pm$.0034
    & .8070$\pm$.0007 & .5269$\pm$.0018
    & .8066$\pm$.0007 & .5282$\pm$.0026
    & .8055$\pm$.0008 & .5288$\pm$.0007 \\
    & AutoDim
    & .8033$\pm$.0011 & .5297$\pm$.0021
    & .8039$\pm$.0010 & .5308$\pm$.0011
    & .8039$\pm$.0015 & .5314$\pm$.0029
    & .8043$\pm$.0011 & .5300$\pm$.0014
    & .7955$\pm$.0038 & .5435$\pm$.0019 \\
    & ES
    & .8159$\pm$.0005 & .5200$\pm$.0010
    & .8164$\pm$.0002 & .5178$\pm$.0002
    & .8156$\pm$.0004 & .5180$\pm$.0016
    & .8162$\pm$.0005 & .5191$\pm$.0012
    & .8152$\pm$.0005 & .5215$\pm$.0014 \\
    & AutoDenoise
    & .8145$\pm$.0001 & .5193$\pm$.0024
    & .8150$\pm$.0002 & .5200$\pm$.0004
    & .8148$\pm$.0003 & .5199$\pm$.0021
    & .8145$\pm$.0004 & .5193$\pm$.0014
    & .8135$\pm$.0005 & .5228$\pm$.0011 \\
    & DS
    & .8145$\pm$.0003 & .5171$\pm$.0003
    & .8159$\pm$.0005 & .5160$\pm$.0010
    & .8149$\pm$.0007 & .5167$\pm$.0008
    & .8149$\pm$.0004 & .5163$\pm$.0012
    & .8153$\pm$.0005 & .5214$\pm$.0031 \\
    & IDC
    & \underline{.8174}$\pm$.0002 & \underline{.5143}$\pm$.0004
    & \underline{.8177}$\pm$.0004 & \underline{.5141}$\pm$.0004
    & \underline{.8176}$\pm$.0002 & \underline{.5148}$\pm$.0009
    & \underline{.8174}$\pm$.0003 & \underline{.5148}$\pm$.0009
    & \underline{.8164}$\pm$.0003 & \underline{.5192}$\pm$.0018 \\
    & OptIDC
    & \textbf{.8181}$^{*}$$\pm$.0002 & \textbf{.5137}$^{*}$$\pm$.0001
    & \textbf{.8185}$^{*}$$\pm$.0003 & \textbf{.5136}$\pm$.0004
    & \textbf{.8184}$^{*}$$\pm$.0004 & \textbf{.5136}$^{*}$$\pm$.0005
    & \textbf{.8182}$^{*}$$\pm$.0003 & \textbf{.5142}$\pm$.0007
    & \textbf{.8172}$^{*}$$\pm$.0003 & \textbf{.5172}$\pm$.0009 \\
    
    \midrule
    \multirow{12}{*}{\rotatebox{90}{AliCCP}}
    & Backbone
    & .6243$\pm$.0001 & \underline{.1616}$\pm$.0000
    & .6241$\pm$.0003 & .1618$\pm$.0001
    & .6238$\pm$.0005 & \underline{.1615}$\pm$.0000
    & .6241$\pm$.0002 & \underline{.1616}$\pm$.0000
    & .6247$\pm$.0003 & \textbf{.1616}$\pm$.0001 \\
    & AutoField
    & .6236$\pm$.0006 & .1630$\pm$.0005
    & .6228$\pm$.0006 & .1629$\pm$.0004
    & .6235$\pm$.0007 & .1629$\pm$.0005
    & .6233$\pm$.0005 & .1629$\pm$.0005
    & .6236$\pm$.0004 & .1625$\pm$.0004 \\
    & LPFS
    & .6244$\pm$.0004 & .1623$\pm$.0007
    & .6247$\pm$.0003 & .1623$\pm$.0003
    & .6249$\pm$.0003 & .1619$\pm$.0003
    & .6237$\pm$.0005 & .1628$\pm$.0002
    & .6240$\pm$.0008 & .1627$\pm$.0003 \\
    & OptEmbed
    & .6242$\pm$.0002 & .1625$\pm$.0001
    & .6246$\pm$.0004 & .1623$\pm$.0003
    & .6246$\pm$.0005 & .1617$\pm$.0001
    & .6247$\pm$.0006 & .1618$\pm$.0002
    & .6243$\pm$.0003 & .1622$\pm$.0002 \\
    & OptFS
    & .6245$\pm$.0005 & .1625$\pm$.0002
    & .6247$\pm$.0003 & .1622$\pm$.0001
    & .6243$\pm$.0006 & .1621$\pm$.0002
    & .6240$\pm$.0005 & .1626$\pm$.0003
    & .6246$\pm$.0003 & .1620$\pm$.0001 \\
    & PEP
    & .6233$\pm$.0004 & \textbf{.1615}$\pm$.0002
    & .6232$\pm$.0007 & \textbf{.1616}$\pm$.0001
    & .6240$\pm$.0004 & \textbf{.1613}$\pm$.0001
    & .6237$\pm$.0008 & \textbf{.1615}$\pm$.0001
    & .6229$\pm$.0002 & \textbf{.1616}$\pm$.0001 \\
    & AutoDim
    & .6183$\pm$.0018 & .1628$\pm$.0007
    & .6186$\pm$.0018 & .1627$\pm$.0014
    & .6183$\pm$.0049 & .1625$\pm$.0007
    & .6191$\pm$.0014 & .1631$\pm$.0008
    & .6193$\pm$.0015 & .1624$\pm$.0001 \\
    & ES
    & .6246$\pm$.0004 & .1622$\pm$.0004
    & .6241$\pm$.0004 & .1620$\pm$.0001
    & .6240$\pm$.0007 & .1618$\pm$.0003
    & .6242$\pm$.0005 & .1618$\pm$.0000
    & .6240$\pm$.0006 & .1620$\pm$.0001 \\
    & AutoDenoise
    & .6196$\pm$.0039 & .1643$\pm$.0018
    & .6199$\pm$.0018 & .1631$\pm$.0003
    & .6214$\pm$.0021 & .1637$\pm$.0011
    & .6212$\pm$.0027 & .1641$\pm$.0019
    & .6207$\pm$.0019 & .1635$\pm$.0005 \\
    & DS
    & .6238$\pm$.0005 & .1619$\pm$.0002
    & .6238$\pm$.0008 & \underline{.1617}$\pm$.0003
    & .6234$\pm$.0007 & .1616$\pm$.0002
    & .6242$\pm$.0005 & .1620$\pm$.0001
    & .6240$\pm$.0001 & \underline{.1617}$\pm$.0002 \\
    & IDC
    & \underline{.6256}$\pm$.0003 & .1621$\pm$.0001
    & \underline{.6254}$\pm$.0004 & .1619$\pm$.0002
    & \underline{.6255}$\pm$.0002 & .1618$\pm$.0000
    & \underline{.6253}$\pm$.0004 & .1623$\pm$.0002
    & \underline{.6256}$\pm$.0003 & .1620$\pm$.0001 \\
    & OptIDC
    & \textbf{.6263}$^{*}$$\pm$.0002 & .1619$\pm$.0001
    & \textbf{.6262}$^{*}$$\pm$.0003 & .1618$\pm$.0000
    & \textbf{.6261}$^{*}$$\pm$.0002 & .1616$\pm$.0001
    & \textbf{.6264}$^{*}$$\pm$.0003 & .1620$\pm$.0002
    & \textbf{.6264}$^{*}$$\pm$.0002 & .1618$\pm$.0001 \\
    \specialrule{0.1em}{1pt}{1pt}
\end{tabular}}
\label{tab:appendix:main_result}
\end{table*}

\subsection{Ablation Study}\label{sec:appendix:ablation}
In this subsection, we provide additional ablation results on the Criteo dataset in Table~\ref{tab:appendix:ablation_result}, using the same variants and the AOP baseline defined in Section~\ref{sec:experiments:rq2}.
Joint configuration generally improves upon single-level selection, although IDC does not consistently outperform the two-level variants. Adding OT coordination to the two-level variants improves or maintains both metrics across all five backbones at the reported precision. For the three-level configuration, OptIDC improves both AUC and LogLoss over IDC across all five backbones.
Removing any individual OT term lowers AUC across all five backbones, while LogLoss increases or remains unchanged. OptIDC also outperforms AOP in both metrics across all five backbones. These results further support the effectiveness of joint configuration and the contribution of each cross-level coordination relation.

\subsection{Details of Coordination Alternatives}
\label{sec:appendix:coordination-alternatives}
We detail the alternatives in Section~\ref{subsec:indepth}. All alternatives sum the three relation penalties and average over samples as in Eq.~\eqref{eq:coordination-target}.
For each alternative, $\lambda_{\mathrm{co}}$ is independently tuned over the same range as OptIDC (Section~\ref{sec:implementation}), with the remaining objective terms unchanged.

\noindent\textbf{w WED and w EED.}
These variants replace only the ground costs in Eq.~\eqref{eq:three-transport-costs}, retaining its selection masses,
temperature-normalized allocations, and averaging factors.
For w WED, the costs are distances between scalar selection weights:
\begin{equation}
\begin{aligned}
c_{ij}^{(\mathrm{WED},\mathrm{sf})}&=|s_i-g_{ij}|, \quad
c_{ijk}^{(\mathrm{WED},\mathrm{fe})}&=|g_{ij}-M_{ijk}|,\\
c_{ijk}^{(\mathrm{WED},\mathrm{se})}&=|s_i-M_{ijk}|.
\end{aligned}
\label{eq:appendix:wed-costs}
\end{equation}
For w EED, the cost uses the same representation pairs as OptIDC, with non-squared Euclidean distance after unit-L2 normalization:
\begin{equation}
c_{\mathrm{EED}}(\mathbf a,\mathbf b)
=\left\|\frac{\mathbf a}{\|\mathbf a\|_2}
       -\frac{\mathbf b}{\|\mathbf b\|_2}\right\|_2,
\label{eq:appendix:normalized-euclidean}
\end{equation}
where $\mathbf a$ and $\mathbf b$ denote the nonzero representations being compared. The representation pairs are $(\overline{\mathbf h}_i,\mathbf h_{ij})$, $(\mathbf h_{ij},\mathbf u_{jk})$, and $(\overline{\mathbf h}_i,\mathbf u_{jk})$ for the sample--feature, feature--unit, and sample--unit relations, respectively.

\noindent\textbf{w CS and w KL.}
For $r\in\{\mathrm{sf},\mathrm{fe},\mathrm{se}\}$, these variants replace $\mathcal W_r(i)$ with a direct comparison of the broadcast selection vectors:
\begin{equation}
\begin{aligned}
(\mathbf a_i^{\mathrm{sf}},\mathbf b_i^{\mathrm{sf}})
  &=(s_i\mathbf 1_F,\mathbf g_i),\\
(\mathbf a_i^{\mathrm{fe}},\mathbf b_i^{\mathrm{fe}})
  &=(\operatorname{vec}(\mathbf g_i\mathbf 1_d^\top),
     \operatorname{vec}(\mathbf M_i)),\\
(\mathbf a_i^{\mathrm{se}},\mathbf b_i^{\mathrm{se}})
  &=(s_i\mathbf 1_{Fd},\operatorname{vec}(\mathbf M_i)).
\end{aligned}
\label{eq:appendix:broadcast-pairs}
\end{equation}
Then, w CS uses the following calculation:
\begin{equation}
\mathcal W_r^{(\mathrm{CS})}(i)
=
\omega_i^r
\left[
1-
\frac{(\mathbf a_i^r)^\top\mathbf b_i^r}
{\max(\|\mathbf a_i^r\|_2,\varepsilon)
 \max(\|\mathbf b_i^r\|_2,\varepsilon)}
\right],
\label{eq:appendix:cs-penalty}
\end{equation}
where $\varepsilon=10^{-8}$, $\omega_i^{\mathrm{sf}}=\omega_i^{\mathrm{se}}=s_i$ and $\omega_i^{\mathrm{fe}}=1$.
For w KL, normalize each complete vector $\mathbf z\in\mathbb R_+^m$ by:
\begin{equation}
\mathcal P_{\varepsilon}(\mathbf z)
=\frac{\mathbf z+\varepsilon\mathbf 1_m}
{\mathbf 1_m^\top\mathbf z+m\varepsilon}.
\label{eq:appendix:kl-normalization}
\end{equation}
Let $\mathbf p_i^r=\mathcal P_{\varepsilon}(\mathbf a_i^r)$ and
$\mathbf q_i^r=\mathcal P_{\varepsilon}(\mathbf b_i^r)$.
The symmetric KL penalty is:
\begin{equation}
\mathcal W_r^{(\mathrm{KL})}(i)
=
\omega_i^r
\sum_l
(p_{il}^r-q_{il}^r)
\log\frac{p_{il}^r}{q_{il}^r}.
\label{eq:appendix:kl-penalty}
\end{equation}

\subsection{Selection Ratio Analysis}\label{subsec:appendix:ratio}
\begin{figure}[h]
    \centering
    \includegraphics[width=0.48\textwidth]{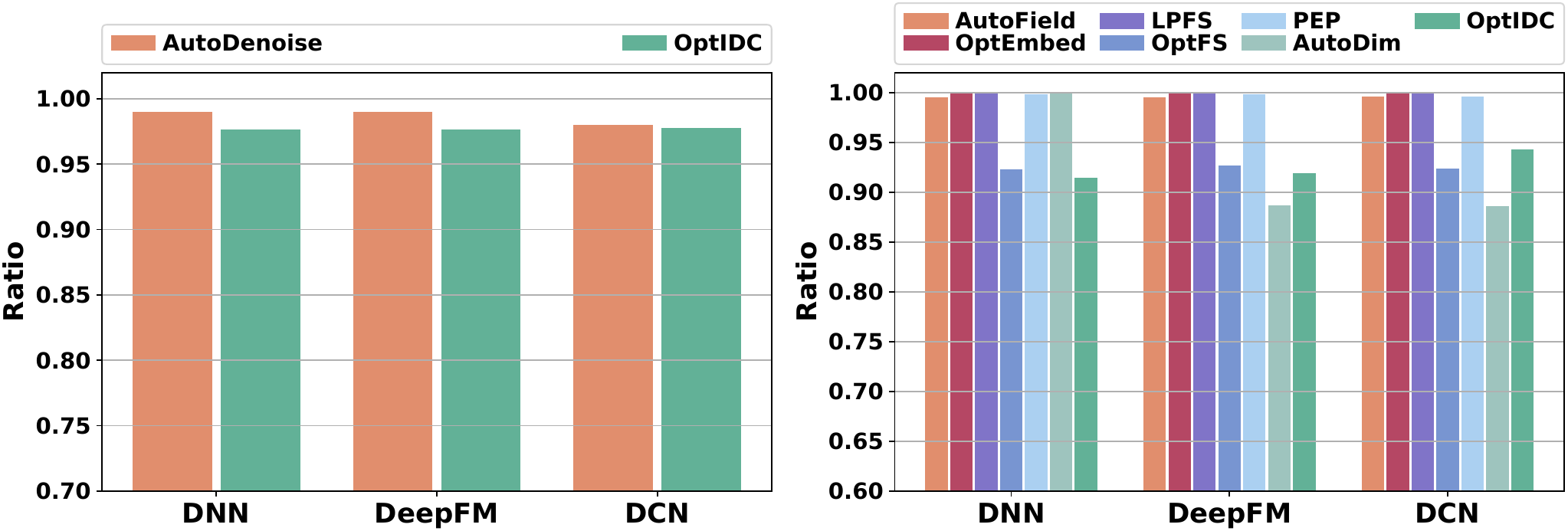}
    \Description[<Figure 7. Fully described in the text.>]{<A full description of Figure 7 can be found in Appendix C.4.>}
    \caption{Comparison of sample retention ratio and embedding retention ratio between OptIDC and baselines on the Avazu dataset.}
    \label{fig:select_ratio}
    \vspace{-10pt}
\end{figure}

To verify that OptIDC can more effectively identify critical input-side information (e.g., high-value samples, informative features, and key embedding units) for achieving more effective input-side dimensional configuration, we compare the sample retention ratio and embedding parameter retention ratio (jointly determined by feature selection and embedding-unit selection) of OptIDC against baseline methods using three backbone models on the Avazu dataset, as shown in Fig.~\ref{fig:select_ratio}. The results show that, for both sample selection and embedding parameter selection, OptIDC consistently achieves superior predictive performance while generally retaining fewer samples and embedding parameters than most baselines. These findings indicate that OptIDC can more precisely identify prediction-relevant information, thereby enabling more efficient input-side dimensional configuration.

\subsection{Hyperparameter Sensitivity Analysis}\label{subsec:appendix:hyper}
\begin{figure*}[htbp]
    \centering
    \includegraphics[width=1.\textwidth]{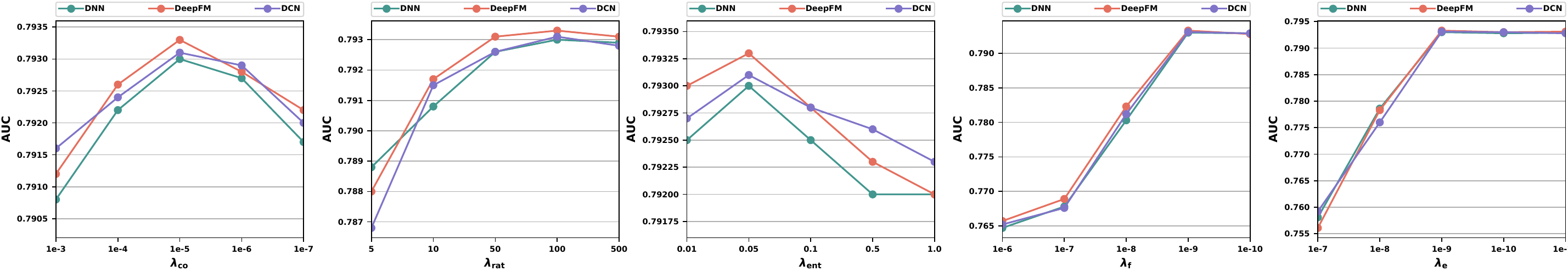}
    \Description[<Figure 8. Fully described in the text.>]{<A full description of Figure 8 can be found in Appendix C.5.>}
    \caption{Sensitivity analysis of key parameters is performed on the Avazu dataset.}
    \label{fig:hyper}
\end{figure*}

Next, we conducted experiments on the Avazu dataset using three backbone models to investigate the sensitivity of five key hyperparameters: (i) $\lambda_{\mathrm{co}}$, (ii) $\lambda_{\mathrm{rat}}$, (iii) $\lambda_{\mathrm{ent}}$, (iv) $\lambda_{\mathrm{f}}$, and (v) $\lambda_{\mathrm{e}}$.
As described in Section~\ref{sec:implementation}, these hyperparameters are varied within their respective predefined ranges.
The results are shown in Fig.~\ref{fig:hyper}, where the y-axis denotes the AUC score and the x-axis represents the search range of each hyperparameter.
From Fig.~\ref{fig:hyper}, we observe that on the Avazu dataset, OptIDC exhibits broadly similar performance trends across the three backbone models. Performance remains relatively stable over parts of the tested ranges, while large feature- and embedding-unit sparsity penalties can substantially degrade AUC. These results suggest that appropriate regularization strengths are important for maintaining predictive performance across different backbones.

\begin{table*}[htbp]
\centering
\caption{Ablation analysis of our method on the Criteo dataset, with the best results marked in bold.}
\resizebox{1.0\textwidth}{!}{
\begin{tabular}{c|c|cccccccccc}
\specialrule{0.1em}{1pt}{1pt}
    \multirow{2}{*}{Dataset}
        &  \multirow{2}{*}{Method}
            & \multicolumn{2}{c|}{DNN} & \multicolumn{2}{c|}{DeepFM} & \multicolumn{2}{c|}{DCN} & \multicolumn{2}{c|}{AutoInt} & \multicolumn{2}{c}{FinalMLP} \\
    \cline{3-12}
    & & AUC$\uparrow$ & LogLoss$\downarrow$ & AUC$\uparrow$ & LogLoss$\downarrow$ & AUC$\uparrow$ & LogLoss$\downarrow$ & AUC$\uparrow$ & LogLoss$\downarrow$ & AUC$\uparrow$ & LogLoss$\downarrow$ \\
    \midrule

    \multirow{12}{*}{\rotatebox{90}{Criteo}}
    & AOP & 0.8106 & 0.4407 & 0.8104 & 0.4409 & 0.8107 & 0.4406 & 0.8103 & 0.4417 & 0.8108 & 0.4410 \\
    & DSFS & 0.8109 & 0.4408 & 0.8106 & 0.4410 & 0.8110 & 0.4406 & 0.8109 & 0.4409 & 0.8109 & 0.4408 \\
    & DSFS+OT & 0.8110 & 0.4406 & 0.8107 & 0.4409 & 0.8112 & 0.4405 & 0.8110 & 0.4408 & 0.8111 & 0.4405 \\
    & DSES & 0.8108 & 0.4407 & 0.8107 & 0.4409 & 0.8109 & 0.4408 & 0.8109 & 0.4406 & 0.8110 & 0.4408 \\
    & DSES+OT & 0.8109 & 0.4407 & 0.8108 & 0.4408 & 0.8110 & 0.4407 & 0.8110 & 0.4405 & 0.8112 & 0.4404 \\
    & FSES & 0.8110 & 0.4406 & 0.8110 & 0.4406 & 0.8110 & 0.4406 & 0.8109 & 0.4408 & 0.8109 & 0.4407 \\
    & FSES+OT & 0.8112 & \textbf{0.4404} & 0.8111 & 0.4405 & 0.8110 & 0.4406 & 0.8110 & 0.4406 & 0.8112 & 0.4403 \\
    & IDC & 0.8110 & 0.4408 & 0.8110 & 0.4407 & 0.8110 & 0.4408 & 0.8110 & 0.4407 & 0.8110 & 0.4407 \\
    & w/o $\mathcal W_{\mathrm{sf}}$ & 0.8111 & 0.4405 & 0.8112 & 0.4406 & 0.8112 & \textbf{0.4404} & 0.8111 & 0.4405 & 0.8114 & 0.4403 \\
    & w/o $\mathcal W_{\mathrm{se}}$ & 0.8111 & \textbf{0.4404} & 0.8112 & 0.4405 & 0.8112 & \textbf{0.4404} & 0.8111 & 0.4406 & 0.8113 & 0.4403 \\
    & w/o $\mathcal W_{\mathrm{fe}}$ & 0.8112 & 0.4406 & 0.8111 & 0.4406 & 0.8112 & 0.4405 & 0.8112 & \textbf{0.4404} & 0.8113 & 0.4403 \\
    & OptIDC & \textbf{0.8113} & \textbf{0.4404} & \textbf{0.8114} & \textbf{0.4404} & \textbf{0.8114} & \textbf{0.4404} & \textbf{0.8113} & \textbf{0.4404} & \textbf{0.8115} & \textbf{0.4402} \\
    
    \specialrule{0.1em}{1pt}{1pt}
\end{tabular}}
\label{tab:appendix:ablation_result}
\end{table*}

%% file: sample-base.bib
@String{Computing = "Computing" }

@String{Computer = "{IEEE} Computer" }

@inproceedings{DeepFM,
  title={DeepFM: A factorization-machine based neural network for CTR prediction},
  author={Guo, Huifeng and Tang, Ruiming and Ye, Yunming and Li, Zhenguo and He, Xiuqiang},
  booktitle={Proceedings of the 26th International Joint Conference on Artificial Intelligence},
  pages={1725--1731},
  year={2017}
}

@inproceedings{autofield,
  title={Autofield: Automating feature selection in deep recommender systems},
  author={Wang, Yejing and Zhao, Xiangyu and Xu, Tong and Wu, Xian},
  booktitle={Proceedings of the ACM Web Conference 2022},
  pages={1977--1986},
  year={2022}
}

@inproceedings{optfs,
  title={Optimizing feature set for click-through rate prediction},
  author={Lyu, Fuyuan and Tang, Xing and Liu, Dugang and Chen, Liang and He, Xiuqiang and Liu, Xue},
  booktitle={Proceedings of the ACM Web Conference 2023},
  pages={3386--3395},
  year={2023}
}

@inproceedings{optembed,
  title={OptEmbed: Learning optimal embedding table for click-through rate prediction},
  author={Lyu, Fuyuan and Tang, Xing and Zhu, Hong and Guo, Huifeng and Zhang, Yingxue and Tang, Ruiming and Liu, Xue},
  booktitle={Proceedings of the 31st ACM International Conference on Information and Knowledge Management},
  pages={1399--1409},
  year={2022}
}

@article{LPFS,
  title={LPFS: Learnable polarizing feature selection for click-through rate prediction},
  author={Guo, Yi and Liu, Zhaocheng and Tan, Jianchao and Liao, Chao and Yang, Sen and Yuan, Lei and Kong, Dongying and Chen, Zhi and Liu, Ji},
  journal={arXiv preprint arXiv:2206.00267},
  year={2022}
}

@inproceedings{AutoDim,
  title={Autodim: Field-aware embedding dimension search in recommender systems},
  author={Zhao, Xiangyu and Liu, Haochen and Liu, Hui and Tang, Jiliang and Guo, Weiwei and Shi, Jun and Wang, Sida and Gao, Huiji and Long, Bo},
  booktitle={Proceedings of the Web Conference 2021},
  pages={3015--3022},
  year={2021}
}

@inproceedings{PEP,
  title={Learnable embedding sizes for recommender systems},
  author={Liu, Siyi and Gao, Chen and Chen, Yihong and Jin, Depeng and Li, Yong},
  booktitle={Proceedings of the 9th International Conference on Learning Representations},
  year={2021}
}

@inproceedings{wang2017deep,
  title={Deep \& cross network for ad click predictions},
  author={Wang, Ruoxi and Fu, Bin and Fu, Gang and Wang, Mingliang},
  booktitle={Proceedings of the ADKDD 2017},
  pages={1--7},
  year={2017}
}

@inproceedings{Autodenoise,
  title={Autodenoise: Automatic data instance denoising for recommendations},
  author={Lin, Weilin and Zhao, Xiangyu and Wang, Yejing and Zhu, Yuanshao and Wang, Wanyu},
  booktitle={Proceedings of the ACM Web Conference 2023},
  pages={1003--1011},
  year={2023}
}

@inproceedings{erase,
  title={Erase: Benchmarking feature selection methods for deep recommender systems},
  author={Jia, Pengyue and Wang, Yejing and Du, Zhaocheng and Zhao, Xiangyu and Wang, Yichao and Chen, Bo and Wang, Wanyu and Guo, Huifeng and Tang, Ruiming},
  booktitle={Proceedings of the 30th ACM SIGKDD Conference on Knowledge Discovery and Data Mining},
  pages={5194--5205},
  year={2024}
}

@inproceedings{MultiFS,
  title={MultiFS: Automated multi-scenario feature selection in deep recommender systems},
  author={Liu, Dugang and Yang, Chaohua and Tang, Xing and Wang, Yejing and Lyu, Fuyuan and Luo, Weihong and He, Xiuqiang and Ming, Zhong and Zhao, Xiangyu},
  booktitle={Proceedings of the 17th ACM International Conference on Web Search and Data Mining},
  pages={434--442},
  year={2024}
}

@inproceedings{SSIM,
  title={Scenario shared instance modeling for click-through rate prediction},
  author={Liu, Dugang and Yang, Chaohua and Fu, Yuwen and Tang, Xing and Li, Gongfu and Lyu, Fuyuan and He, Xiuqiang and Ming, Zhong},
  booktitle={Proceedings of the 31st ACM SIGKDD Conference on Knowledge Discovery and Data Mining},
  pages={2436--2447},
  year={2025}
}

@article{zheng2023automl,
  title={Automl for deep recommender systems: A survey},
  author={Zheng, Ruiqi and Qu, Liang and Cui, Bin and Shi, Yuhui and Yin, Hongzhi},
  journal={ACM Transactions on Information Systems},
  volume={41},
  number={4},
  pages={1--38},
  year={2023}
}

@inproceedings{lin2022adafs,
  title={AdaFS: Adaptive feature selection in deep recommender system},
  author={Lin, Weilin and Zhao, Xiangyu and Wang, Yejing and Xu, Tong and Wu, Xian},
  booktitle={Proceedings of the 28th ACM SIGKDD International Conference on Knowledge Discovery and Data Mining},
  pages={3309--3317},
  year={2022}
}

@inproceedings{liu2020automated,
  title={Automated embedding size search in deep recommender systems},
  author={Liu, Haochen and Zhao, Xiangyu and Wang, Chong and Liu, Xiaobing and Tang, Jiliang},
  booktitle={Proceedings of the 43rd International ACM SIGIR Conference on Research and Development in Information Retrieval},
  pages={2307--2316},
  year={2020}
}

@inproceedings{liu2020autofis,
  title={Autofis: Automatic feature interaction selection in factorization models for click-through rate prediction},
  author={Liu, Bin and Zhu, Chenxu and Li, Guilin and Zhang, Weinan and Lai, Jincai and Tang, Ruiming and He, Xiuqiang and Li, Zhenguo and Yu, Yong},
  booktitle={proceedings of the 26th ACM SIGKDD International Conference on Knowledge Discovery and Data Mining},
  pages={2636--2645},
  year={2020}
}

@inproceedings{lyu2023towards,
  title={Towards hybrid-grained feature interaction selection for deep sparse network},
  author={Lyu, Fuyuan and Tang, Xing and Liu, Dugang and Ma, Chen and Luo, Weihong and Chen, Liang and He, Xiuqiang and Liu, Xue},
  booktitle={Proceedings of the 37th Conference on Neural Information Processing Systems},
  pages={49325--49340},
  year={2023}
}

@inproceedings{zhao2021autoloss,
  title={AutoLoss: Automated loss function search in recommendations},
  author={Zhao, Xiangyu and Liu, Haochen and Fan, Wenqi and Liu, Hui and Tang, Jiliang and Wang, Chong},
  booktitle={Proceedings of the 27th ACM SIGKDD International Conference on Knowledge Discovery and Data Mining},
  pages={3959--3967},
  year={2021}
}

@inproceedings{chen2019lambdaopt,
  title={$\lambda$opt: Learn to regularize recommender models in finer levels},
  author={Chen, Yihong and Chen, Bei and He, Xiangnan and Gao, Chen and Li, Yong and Lou, Jian-Guang and Wang, Yue},
  booktitle={Proceedings of the 25th ACM SIGKDD International Conference on Knowledge Discovery and Data Mining},
  pages={978--986},
  year={2019}
}

@inproceedings{gao2023autotransfer,
  title={AutoTransfer: Instance transfer for cross-domain recommendations},
  author={Gao, Jingtong and Zhao, Xiangyu and Chen, Bo and Yan, Fan and Guo, Huifeng and Tang, Ruiming},
  booktitle={Proceedings of the 46th International ACM SIGIR Conference on Research and Development in Information Retrieval},
  pages={1478--1487},
  year={2023}
}

@inproceedings{thanthriwatta2021instance,
  title={Instance selection for online updating in dynamic recommender environments},
  author={Thanthriwatta, Thilina and Rosenblum, David S},
  booktitle={Proceedings of the 25th Pacific-Asia Conference on Knowledge Discovery and Data Mining},
  pages={612--624},
  year={2021}
}

@inproceedings{song2019autoint,
  title={Autoint: Automatic feature interaction learning via self-attentive neural networks},
  author={Song, Weiping and Shi, Chence and Xiao, Zhiping and Duan, Zhijian and Xu, Yewen and Zhang, Ming and Tang, Jian},
  booktitle={Proceedings of the 28th ACM International Conference on Information and Knowledge Management},
  pages={1161--1170},
  year={2019}
}

@inproceedings{rendle2010fm,
  title={Factorization machines},
  author={Rendle, Steffen},
  booktitle={Proceedings of the 2010 IEEE International Conference on Data Mining},
  pages={995--1000},
  year={2010}
}

@inproceedings{cheng2016wide,
  title={Wide \& deep learning for recommender systems},
  author={Cheng, Heng-Tze and Koc, Levent and Harmsen, Jeremiah and Shaked, Tal and Chandra, Tushar and Aradhye, Hrishi and Anderson, Glen and Corrado, Greg and Chai, Wei and Ispir, Mustafa and others},
  booktitle={Proceedings of the 1st Workshop on Deep Learning for Recommender Systems},
  pages={7--10},
  year={2016}
}

@inproceedings{ma2018modeling,
  title={Modeling task relationships in multi-task learning with multi-gate mixture-of-experts},
  author={Ma, Jiaqi and Zhao, Zhe and Yi, Xinyang and Chen, Jilin and Hong, Lichan and Chi, Ed H},
  booktitle={Proceedings of the 24th ACM SIGKDD International Conference on Knowledge Discovery and Data Mining},
  pages={1930--1939},
  year={2018}
}

@inproceedings{ple,
  title={Progressive layered extraction (ple): A novel multi-task learning (mtl) model for personalized recommendations},
  author={Tang, Hongyan and Liu, Junning and Zhao, Ming and Gong, Xudong},
  booktitle={Proceedings of the 14th ACM Conference on Recommender Systems},
  pages={269--278},
  year={2020}
}

@inproceedings{cuturi2013sinkhorn,
  title={Sinkhorn distances: Lightspeed computation of optimal transport},
  author={Cuturi, Marco},
  booktitle = {Proceedings of the 27th International Conference on Neural Information Processing Systems},
  year={2013}
}

@article{zhang2019deep,
  title={Deep learning based recommender system: A survey and new perspectives},
  author={Zhang, Shuai and Yao, Lina and Sun, Aixin and Tay, Yi},
  journal={ACM Computing Surveys},
  volume={52},
  number={1},
  pages={1--38},
  year={2019}
}

@inproceedings{wang2025universal,
  title={A universal framework for compressing embeddings in ctr prediction},
  author={Wang, Kefan and Wang, Hao and Song, Kenan and Guo, Wei and Cheng, Kai and Li, Zhi and Liu, Yong and Lian, Defu and Chen, Enhong},
  booktitle={International Conference on Database Systems for Advanced Applications},
  pages={84--100},
  year={2025}
}

@inproceedings{finalmlp,
  title={FinalMLP: An enhanced two-stream MLP model for CTR prediction},
  author={Mao, Kelong and Zhu, Jieming and Su, Liangcai and Cai, Guohao and Li, Yuru and Dong, Zhenhua},
  booktitle={Proceedings of the AAAI Conference on Artificial Intelligence},
  pages={4552--4560},
  year={2023}
}

@article{choi2025overcoming,
  title={Overcoming spurious solutions in semi-dual neural optimal transport: A smoothing approach for learning the optimal transport plan},
  author={Choi, Jaemoo and Choi, Jaewoong and Kwon, Dohyun},
  journal={arXiv preprint arXiv:2502.04583},
  year={2025}
}

@inproceedings{ge2021ota,
  title={Ota: Optimal transport assignment for object detection},
  author={Ge, Zheng and Liu, Songtao and Li, Zeming and Yoshie, Osamu and Sun, Jian},
  booktitle={Proceedings of the IEEE/CVF conference on computer vision and pattern recognition},
  pages={303--312},
  year={2021}
}

@inproceedings{yin2025feature,
  title={From feature interaction to feature generation: A generative paradigm of CTR prediction models},
  author={Yin, Mingjia and Pan, Junwei and Wang, Hao and Wang, Ximei and Zhang, Shangyu and Jiang, Jie and Lian, Defu and Chen, Enhong},
  booktitle={International Conference on Machine Learning},
  pages={72393--72413},
  year={2025}
}
